\documentclass[aps,prl,twocolumn,superscriptaddress,showpacs,floatfix,maxbibnames=10]{revtex4-2}

\usepackage{amsmath,amsthm,amssymb,amsfonts,float,graphics,epsfig,epstopdf,color,verbatim,tabularx,bm,multirow,appendix}
\usepackage[utf8]{inputenc}
\usepackage[T1]{fontenc}
\usepackage{xcolor}
\usepackage{dsfont}
\usepackage{textcomp}
\usepackage{yfonts}
\usepackage{footnote}
\usepackage{bm}
\usepackage{subfigure}
\usepackage{mathrsfs}
\usepackage{MnSymbol}
\usepackage{mathtools}
\usepackage{graphicx}
\usepackage{verbatim}
\usepackage[colorlinks=true, citecolor=blue, linkcolor=blue, urlcolor=blue]{hyperref}
\usepackage{multirow}
\usepackage{braket}
\usepackage[normalem]{ulem}
\usepackage{tikz}
\usetikzlibrary{calc}
\usetikzlibrary{shapes.multipart}
\usepackage{orcidlink}
\usepackage{xr} 
\usepackage[percent]{overpic}

\newcommand{\dist}{d}


\begin{document}

\title{Taming multiparty entanglement at measurement-induced phase transitions}

\author{Liuke Lyu}
\altaffiliation{The two authors contributed equally to this work.}
\affiliation{D\'epartement de Physique, Universit\'e de Montr\'eal, Montr\'eal, QC H3C 3J7, Canada}
\affiliation{
 Institut Courtois, Universit\'e de Montr\'eal, Montr\'eal (Qu\'ebec), H2V 0B3, Canada
}
\affiliation{
 Centre de Recherches Math\'ematiques, Universit\'e de Montr\'eal, Montr\'eal, QC, Canada, HC3 3J7
}

\author{James Allen}
\altaffiliation{The two authors contributed equally to this work.}
\affiliation{D\'epartement de Physique, Universit\'e de Montr\'eal, Montr\'eal, QC H3C 3J7, Canada}

\author{Yi Hong Teoh}
\affiliation{Department of Physics and Astronomy, University of Waterloo, Ontario, N2L 3G1, Canada}

\author{Roger G Melko}
\affiliation{Department of Physics and Astronomy, University of Waterloo, Ontario, N2L 3G1, Canada}
\affiliation{Perimeter Institute for Theoretical Physics, Waterloo, Ontario, N2L 2Y5, Canada}

\author{William Witczak-Krempa}
\email{w.witczak-krempa@umontreal.ca}
\affiliation{D\'epartement de Physique, Universit\'e de Montr\'eal, Montr\'eal, QC H3C 3J7, Canada}
\affiliation{Institut Courtois, Universit\'e de Montr\'eal, Montr\'eal (Qu\'ebec), H2V 0B3, Canada}
\affiliation{ Centre de Recherches Math\'ematiques, Universit\'e de Montr\'eal, Montr\'eal, QC, Canada, HC3 3J7}

\date{\today}

\begin{abstract}
Measurement-induced phase transitions (MIPT) give rise to novel dynamical states of quantum matter realized by balancing unitary evolution and measurements. We present large-scale numerical simulations of a trapped-ion native MIPT, argued to belong to the universality class described by the Haar non-unitary conformal field theory. First, through a finite-size analysis we obtained the critical measurement rate, and correlation length exponent, which falls close to the percolation value. Second, by leveraging a monotone computable via semi-definite programming, we uncover robust algebraic decay of genuine multiparty entanglement (GME) versus separation for 2, 3, and 4 parties. The corresponding critical exponents are lower-bounded by those of the multiparty mutual information, which we determine up to 4 parties, and conjecture to be (k+2) for k parties. Additionally, we derive lower bounds for both GME and multiparty mutual information. 
\end{abstract}

\maketitle

\section{Introduction}

\begin{figure}
    \centering
    \includegraphics[width=0.95\linewidth]{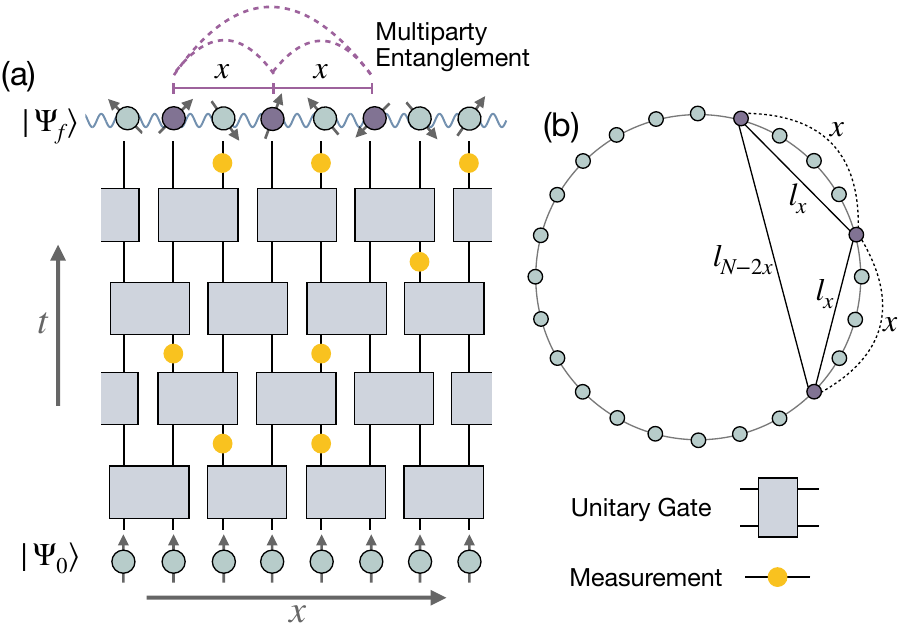}
    \caption{\textbf{Monitored circuit model and conformal geometry.} (a) Space-time diagram of the hybrid circuit dynamics. The system evolves via layers of unitary gates (blocks) applied in a brickwork pattern with periodic boundary conditions, interspersed with single-site projective measurements (yellow circles) performed with probability $p$. We probe the spatial structure of the critical state by computing multiparty entanglement for a subsystem of qubits (highlighted in purple) separated by a distance $x$. (b) Geometric representation of the circuit with periodic boundary conditions. To extract universal scaling exponents, Euclidean distances $x$ are replaced by chord lengths $l_x$ defined by (\ref{eq:chord_length_definition}). }
    \label{fig:circuit}
\end{figure}

Measurement-induced phase transitions (MIPT) constitute novel dynamical quantum critical phase transitions that occur in monitored quantum many-body systems~\cite{Heyl_2018,Skinner2019,Li2019_MIPT,Jian2020_MIPT,Vasseur2019,Pixley2020,RQC}. In one of the most studied realizations, a chain of qubits ($s = 1/2$ spins) is evolved with a brickwork of local two-site unitary operations (Fig.~\ref{fig:circuit}). Between the unitary layers, single-site projective measurements take place with a probability or rate $p$. This leads to an ensemble of quantum states due to the random content of each unitary, the random positions of the measurements, and the measurement outcomes. When $p$ is large, the measurements keep the state close to a product state: this is the low-entanglement area-law regime. When $p$ is low, the unitaries rapidly generate global entanglement among all the qubits (scrambling)~\cite{Bera2020,Schuster2025,LaRacuente2025}, resulting in a high-entanglement volume-law regime. It was found that an intermediate value $p_c$ leads to critical transition in this ensemble where the entanglement entropy of an interval of size $N_A$ shows a logarithmic growth $\log(N_A)$, like in regular quantum critical points in equilibrium (for instance, the transition in the transverse field Ising model). However, in contrast to the unitary conformal field theory (CFT) description of the equilibrium transitions, MIPTs have been argued~\cite{Skinner2019,Vasseur2019,Jian2020_MIPT,Li2021_MIPT,Zabalo2022,Aziz2024} to be described by non-unitary CFTs~\cite{Gurarie1993,Vasseur2012,Cardy2013,Hogervorst2017}, which are far less understood. In particular, MIPTs show a richer entanglement structure if one looks beyond the entanglement entropy. 

Ground or thermal states of local Hamiltonians, including those at a quantum critical point, have been argued to lack entanglement between sufficiently separated parties~\cite{parez2024fate}. In the simplest case, 2 disjoint groups of spins will lose entanglement beyond a certain separation. This leads to a ``sudden death’’~\cite{Osterloh2002,Osborne2002} that can be observed via the entanglement negativity, for instance~\cite{Javanmard2018,parez2024fate,Wang2025EntMicro}. In contrast, the negativity has been found to decay algebraically with separation in certain MIPTs~\cite{Shi2020,Sang2021MIPTNeg}.
Later, numerical evidence for algebraic decay of the genuine multiparty entanglement (GME) between 3 parties was obtained in the Haar-random MIPT~\cite{Sebastien2025}. Heuristically, GME, which we define explicitly below, arises when all parties participate in the entanglement. 
Aside from the Haar-random system, in an MIPT that maps to classical percolation, GME was found to decay algebraically with separation with an exponent $\alpha_k = 2k$, where $k$ is the number of parties~\cite{allen2025spatial}. A general framework to understand entanglement near MIPTs was put forward based on entanglement clusters. For instance, these have been used to argue for universal entanglement inequalities, such as monotonicity $\alpha_{k+1} \geq \alpha_k$, and subadditivity $\alpha_{k+\ell} \leq \alpha_k +\alpha_\ell$. However, these have not been tested in a generic MIPT, such as the paradigmatic Haar MIPT. 

In this work, we tackle the important question of determining the entanglement exponents $\alpha_k$ by performing large-scale simulations of an ensemble of quantum circuits that can be implemented in trapped-ion architectures~\cite{Sycamore,Pogorelov2021MMS,Czischek2021,hu2025neural}. Compared to the Haar-random circuit ensemble, the current gate set has the advantage of being finite, while maintaining universality in the quantum computation sense. In order to study the GME, we evaluate the genuine multiparty negativity (GMN), a comprehensive entanglement monotone (i.e. nonincreasing under local operations and classical communication, or LOCC) that leverages the power of semidefinite programming (SDP)~\cite{Jungnitsch2011,Hofmann2014}. As such the GMN is one of the rare GME monotones that is easily and reliably computable. In contrast, quantities that are based on regular optimization, such as those used in Ref.~\cite{Sebastien2025}, are too challenging to consistently evaluate, especially in the regime of low GME at large separations. 

By working with chains up to 24 sites, we extract the GME exponents for 2, 3, and 4 parties (Table~\ref{tab:entanglement_exponents}). Finite-size scaling yields estimates in the thermodynamic limit. First, we find a much larger bipartite negativity exponent than previous works: $\alpha_2\approx 9$. Second, our exponents obey the inequalities alluded to above, including monotonicity and subadditivity. Furthermore, we study the scaling of the multiparty mutual information, which includes both classical and entangling correlations, versus separation in chains containing up to 26 sites. Our finite-size extrapolation yields exponents for $k\leq 4$ parties consistent with the relation $\alpha_k^{\rm MI}=k+2$, which we conjecture to hold for all $k$. By exploiting the emergent properties of the MIPT, we establish a lower bound $\alpha_k^{\rm MI} \geq k$, which is obeyed by our conjectured result. Finally, we study the space-time evolution of multiparty entanglement by numerically evaluating entanglement weighted graphs, allowing to construct the entanglement clusters of Ref.~\cite{allen2025spatial}. This visual representation establishes the importance of measurement halos in creating long-range entanglement.

{\renewcommand{\arraystretch}{1.3}
\setlength{\tabcolsep}{6pt}
\begin{table}[h]
\centering
\begin{tabular}{l|ccccc|c}
\hline
\textit{N} & 18 & 20 & 22 & 24 & 26 &$\infty$\\
\hline \rule{0pt}{0.9\normalbaselineskip}$\alpha_2^{\mathrm{GMN}}$ & 6.79 & 7.36 & 8.44 & 8.83 & & $9\,_{-0.2}^{+1.0}$\\
$\alpha_3^{\mathrm{GMN}}$ & 9.24 & 9.48 & 10.16 & 10.61 & & $11\,_{-0.3}^{+2.0}$\\
$\alpha_4^{\mathrm{GMN}}$ & 9.25 & 9.74 & 10.52 & 11.31 & & >12\\ \hline
\rule{0pt}{0.9\normalbaselineskip}$\alpha_2^{\text{MI}}$  & 3.41 & 3.77 & 3.86 & 3.96 & 3.94 & $4\,_{-0.1}^{+0.2}$\\
$\alpha_3^{\text{MI}}$  & 4.75 & 4.74 & 4.95 & 5.05 & 4.95 & $5\,_{-0.1}^{+0.2}$\\
$\alpha_4^{\text{MI}}$  & 5.32 & 5.47 & 5.71 & 5.94 & 5.96 & $6\,_{-0.1}^{+0.4}$\\
\hline
\end{tabular}
\caption{Scaling exponents $\alpha_k$ for genuine multiparty negativity and multiparty mutual information extracted at $p=0.17$ for various system sizes $N$, with approximate $N\to\infty$ limits where convergence is observed. The extrapolation is shown in Fig.~\ref{fig:exponents_over_N}. }
\label{tab:entanglement_exponents}
\end{table}
}

The rest of the paper is structured as follows. Section~\ref{sec:model_and_methods} introduces the circuit model and methods: Sec.~\ref{sec:trapped_ion_native_circuit} defines the trapped-ion native circuit ensemble, Sec.~\ref{sec:measures_of_multiparty_correlations} introduces the multiparty mutual information and genuine multiparty negativity, Sec.~\ref{sec:critical_point} uses the tripartite mutual information to determine the critical point of the ensemble, and Sec.~\ref{sec:scaling_ansatz_and_conformal_geometry} specifies the distance scale used for conformal finite-size analysis.
Section~\ref{sec:results} presents our main results on the real-space scaling of multiparty entanglement 
and the extracted scaling exponents, followed by a comparison to theoretical constraints in Sec.~\ref{sec:exp_lower_bound} and a spacetime visualization via entanglement-weighted graphs in Sec.~\ref{sec:ent_weight_graphs}. We conclude in Section~\ref{sec:conclusion} with a summary and outlook.

\section{Model and Methods} \label{sec:model_and_methods}

To investigate the universal properties of the MIPT, we require a circuit model that balances two competing requirements: it must be generic enough to capture the Haar universality class, yet structured enough to be numerically tractable and experimentally relevant. We employ a (1+1)-dimensional brickwork circuit model constructed from a small gate set native to trapped-ion quantum processors.

\subsection{Trapped-Ion Native Circuit} \label{sec:trapped_ion_native_circuit}

The circuit consists of layers of two-site entangling gates applied to nearest neighbors in a brickwork geometry, interspersed with layers of single-site projective measurements. 
The unitary gates are constructed from the Mølmer-Sørensen (MS) interaction, the native entangling operation in trapped-ion platforms~\cite{Sorensen1971, Czischek2021, Pogorelov2021MMS, hu2025neural}. Specifically, we utilize gates acting on sites $j, j+1$ of the form:
\begin{equation}
    U_{j, j+1} = \mathcal{M}_{j,j+1}  \; R_j \otimes R_{j+1},
\end{equation}
where $\mathcal{M}$ is the fixed-angle Mølmer-Sørensen gate defined as:
\begin{equation}
    \mathcal{M}_{j,j+1} = \exp\left(-i \frac{\pi}{4} X_j X_{j+1}\right),
\end{equation}
and $R_j, R_{j+1}$ are single-qubit rotations. These rotations are drawn uniformly from the discrete set of $\pi/2$ rotations $R(\hat{n}) = \exp(-i \frac{\pi}{4} \hat{n} \cdot \vec{\sigma})$ about the axes:
\(
    \hat{n} \in \left\{\hat{x}, \hat{y}, (\hat{x}+\hat{y})/\sqrt{2}\right\}.
\)
Following the unitary layer, each qubit is measured in the computational $Z$-basis with probability $p$.

While this gate set is discrete, it forms a universal set for quantum computation~\cite{Sycamore}. Consequently, in the limit of large depth, the ensemble of unitaries generated by this circuit converges to the Haar measure (see Appendix~\ref{app:ensemble_universality}). Since the universality class of the MIPT is determined by the symmetry of the underlying unitary dynamics, this model should belong to the same universality class as the standard Haar-random circuit models widely studied in the literature.
A significant advantage of this specific gate set—beyond its simplicity and immediate experimental realizability—is its numerical stability for scaling analysis. 
Standard Haar-random brickwork circuits often exhibit strong parity oscillations (``even--odd'' effect) in entanglement quantities due to the discrete lattice geometry (see Appendix~\ref{app:compare_haar}).
These oscillations make the precise extraction of critical exponents difficult without extremely large system sizes.
In contrast, our interaction term $X_j X_{j+1}$ possesses a self-commuting structure that naturally suppresses these parity effects. This suppression 
enables us to extract reliable critical exponents for high-order multiparty correlations ($k=3, 4$) compared to standard Haar circuits.

To prepare the ensemble of states, we simulate system sizes ranging from $N=18$ to $N=26$. Each realization is initialized in the product state $\ket{0}^{\otimes N}$ and evolved for a depth proportional to the system size. Specifically, the circuit is applied for $N$ full periods, where a single period consists of an even unitary layer (gates on pairs 1-2, 3-4, ...), a measurement layer, an odd unitary layer (2-3 ... N-1), and a second measurement layer. This results in a total depth of $2N$ unitary layers and $2N$ measurement layers. To mitigate any bias arising from the even-odd effect of the final layer, we apply one additional even unitary layer followed by measurements with a probability of $50\%$ at the end of the evolution. We confirmed that this depth is sufficient to reach the non-equilibrium steady state by comparing these results with simulations performed at a depth of $1.5N$ periods (see Appendix Fig.~\ref{fig:depth_comparison}), finding excellent agreement between the two.

\subsection{Measures of Multiparty Correlations} \label{sec:measures_of_multiparty_correlations}

To characterize the complex structure of the critical state, we employ two complementary information-theoretic probes. We first quantify the total correlations (both classical and quantum) shared among subsystems using the quantum mutual information, and subsequently isolate the strictly non-classical contributions using a rigorous witness for genuine multiparty entanglement.

\paragraph{Multiparty Mutual Information}
We begin with the generalized $k$-party mutual information $I_k$. For a collection of $k$ subsystems $\{A_1, A_2, \dots, A_k\}$, this quantity captures the reciprocal information shared among the $k$-parties. It can be defined~\cite{McGill1954_multipartyMI,Casini2009,Hayden2013_multipartyMI} via an alternating sum of the von Neumann entropies of all possible composite subsystems.
Let $\mathcal{K} = \{1, \dots, k\}$ be the set of indices labeling the regions. The $k$-party mutual information is given by:
\begin{equation}
    I_k(A_1 , \dots , A_k) = \sum_{\substack{T \subseteq \mathcal{K} \\ T \neq \emptyset}} (-1)^{|T|+1} S\left(\bigcup_{i \in T} A_i\right),
\end{equation}
where the sum runs over all non-empty subsets $T$ of the index set $\mathcal{K}$, and $S(\rho)=-\text{tr}(\rho\log_2\rho)$ is the von Neumann entropy of the union of regions specified by $T$. For $k=2$, this recovers the standard mutual information $I_2(A,B) = S_A + S_B - S_{AB}$. 
While bipartite mutual information is always non-negative, higher-order mutual information ($k \ge 3$) can be negative. Since $I_3$ is found to be consistently negative for the circuits studied here (similar to Ref.~\cite{Pixley2020}), in what follows we show the negative tripartite mutual information, $-I_3$. 
In this work, we exploit this measure in two distinct regimes. First, we utilize $I_3$ of macroscopic partitions to precisely locate the critical point, leveraging its ability to cancel boundary-law contributions. Second, we employ the general $I_k$ between spatially separated local subregions to quantify the decay of long-range information in the critical state, serving as a baseline for total correlation against which we contrast genuine multiparty entanglement.

\paragraph{Genuine Multiparty Entanglement.}
While mutual information captures correlations of arbitrary origin, we are specifically interested in isolating collective quantum resources. A state $\rho$ is considered genuinely $k$-party entangled if cannot be expressed as a mixture of states that are separable across a bipartition~\cite{Guhne2005,Guhne2009,Guhne2010}. 
Let $\mathcal{M}$ be the set of all possible bipartitions $m = m_1 \cup m_2$ of the $k$ parties. The set of \textit{biseparable states} is defined as the convex hull of states separable with respect to these fixed cuts:
\begin{equation}
    \rho_{\text{bs}} = \sum_{m \in \mathcal{M}} p_m \sum_j q_{j|m} \, \rho_{m_1}^{j} \otimes \rho_{m_2}^{j} \, ,
\end{equation}
where $\{p_m\}$ and $\{q_{j|m}\}$ are probability distributions. Any state $\rho$ that cannot be written in this form is called genuine multiparty entangled (GME).

To strictly quantify this resource, we employ the Genuine Multiparty Negativity (GMN), denoted $\mathcal{N}_k(\rho)$. This measure is computable via a semidefinite program (SDP) that searches for an optimal entanglement witness $\mathcal{W}$ capable of distinguishing $\rho$ from the set of biseparable states. 
For a collection of $k$ subsystems $\{A_1, A_2, \dots, A_k\}$, 
the GMN is defined as:
\begin{equation}
    \mathcal{N}_k(\rho) = - \min_{\mathcal{W}} \text{Tr}(\rho \mathcal{W}),
\end{equation}
subject to the constraint that $\mathcal{W}$ is fully decomposable with respect to all bipartitions of the $k$ parties. Specifically, for every bipartition $m$ of the parties into two subsets $m_1,m_2$, the witness must satisfy $\mathcal{W} = P_m + Q_m^{T_{m_1}}$, where $0 \preceq (P_m, Q_m) \preceq I$ ($A\succeq B$ means $A-B$ is positive semidefinite), and $T_{m_1}$ denotes the partial transpose with respect to the partition $m_1$~\cite{Peres1996}. 
A non-zero value $\mathcal{N}_k > 0$ serves as a rigorous witness that the state possesses genuine $k$-party entanglement. Furthermore, because GMN is non-increasing under local operations and classical communication (LOCC), it is a true GME  monotone~\cite{Jungnitsch2011}, allowing us to track the scaling of collective quantum resources across the phase transition.
In the bipartite case ($k=2$), this monotone reduces to the standard quantum negativity $\mathcal{N}_2(\rho) = \frac{1}{2}\left(||\rho^{T_A}||_1 - 1\right)$, where $\|X\|_1=\text{tr}\sqrt{X^\dagger X}$ is the trace norm.
Crucially, the negativity between qubits is monogamous~\cite{CKW_negativity2007}. This property highlights a fundamental distinction from mutual information: unlike classical correlations, entanglement is a strictly exclusive resource. Because quantum information cannot be copied, a qubit maximally entangled with one partner cannot share entanglement with any other. This monogamy implies that entanglement cannot be freely distributed among many parties, making the generation of long-range entanglement significantly more difficult than the propagation of classical information.

\subsection{Determining the Critical Point}
\label{sec:critical_point}
Before characterizing the critical state, we first determine the precise location of the phase transition for the MMS gate set, using the tripartite mutual information ($I_3$).
Following Ref.~\cite{Pixley2020}, we partition the system of size $N$ (with periodic boundary conditions) into four equal contiguous intervals $A, B, C, D$. This specific partitioning is chosen because the alternating sum structure of $I_3(A,B,C)$ cancels out the volume and/or boundary-law terms of the entanglement entropy. Consequently, $I_3$ allows for a sharp identification of the critical point. 

\begin{figure}[hbt!]
    \centering
    \includegraphics[width=1\columnwidth]{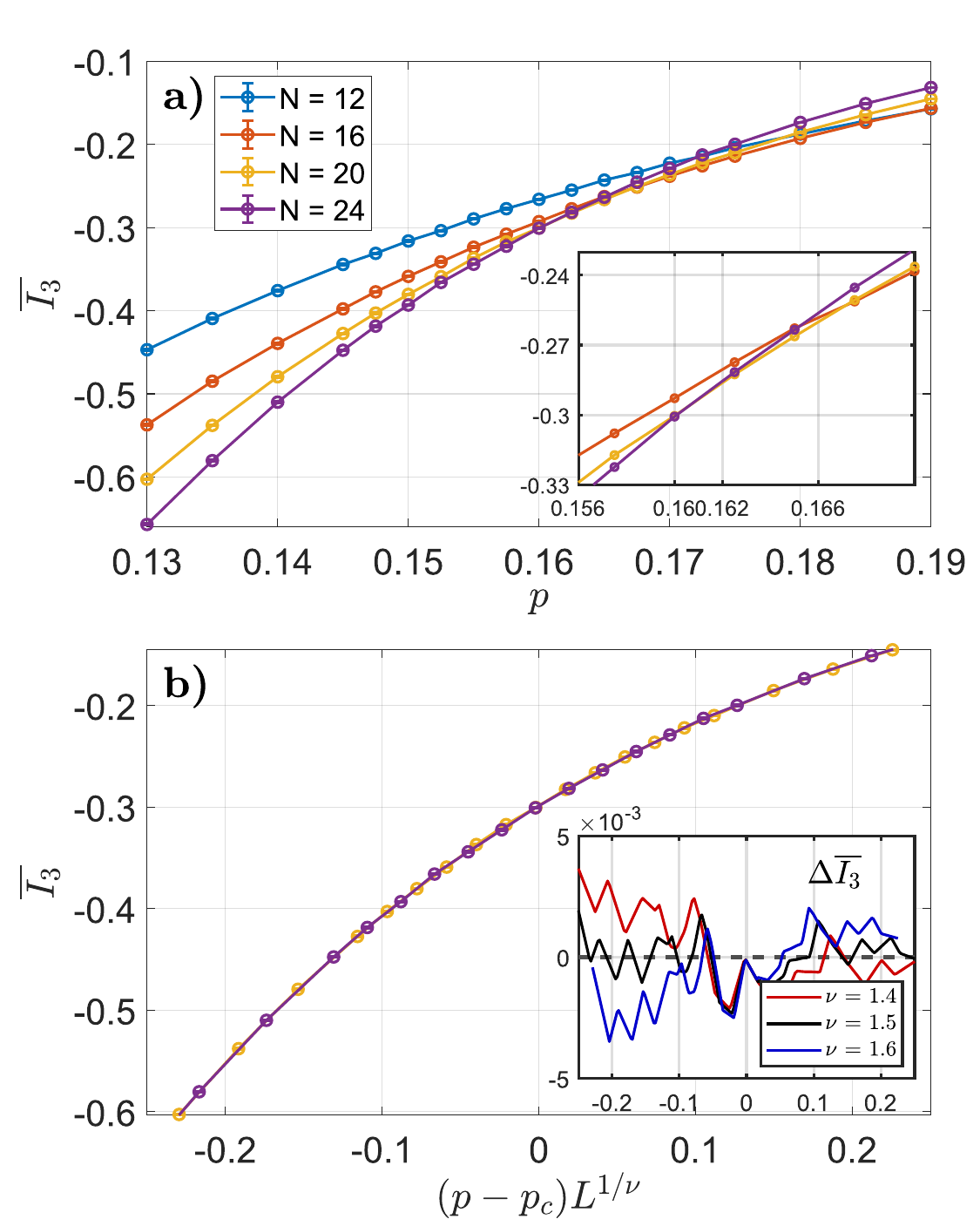}
    \caption{\textbf{Critical point identification via tripartite mutual information (TMI).} \textbf{(a)} TMI plotted as a function of measurement probability $p$ for system sizes $N \in \{12, 16, 20, 24\}$. \textit{Inset:} The intersection of the curves for the largest system sizes, $N=20$ and $N=24$ shows that the critical measurement rate $p_c \lesssim 0.16$. \textbf{(b)} Finite-size scaling collapse of the TMI data. By rescaling the horizontal axis as $(p-p_c)L^{1/\nu}$, the data for $N=20$ and $24$ collapse onto a single universal curve. 
    \textit{Inset:} A collapse-diagnostic based on the difference between these two rescaled curves for different $\nu$.
    Optimizing the collapse over $\nu$ and $p_c$ yields a correlation length critical exponent $\nu = 1.48(4)$ at $p_c = 0.160(1)$ for the specific pair of curves at $N=20, 24$. 
    }
    \label{fig:TMI_melko}
\end{figure}

\begin{figure}[hbt!]
    \centering
    \begin{tikzpicture}
        \begin{scope}
            \node[anchor=north west,inner sep=0] (image_a) at (0,0)
            {\includegraphics[width=0.48\columnwidth]{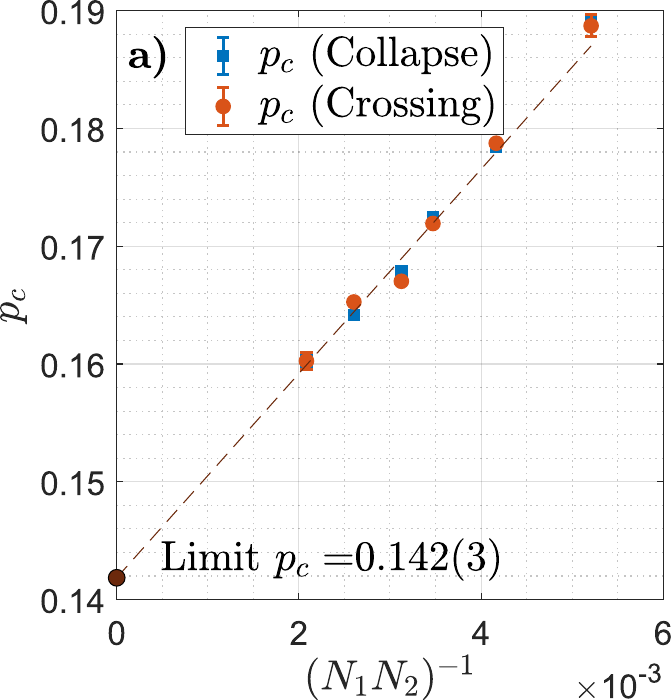}};
        \end{scope}
        \begin{scope}[xshift=0.5\columnwidth]
            \node[anchor=north west,inner sep=0] (image_a) at (0,0)
            {\includegraphics[width=0.48\columnwidth]{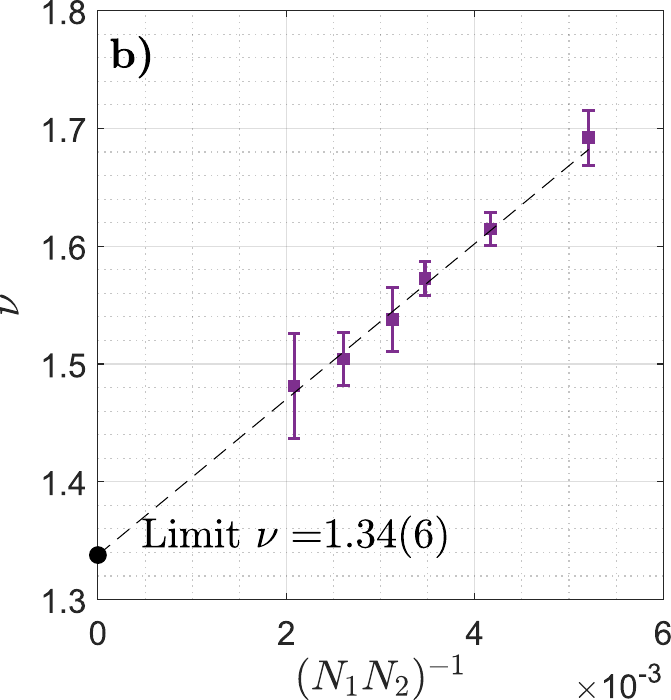}};
        \end{scope}
    \end{tikzpicture}
    \caption{\textbf{Large-N limits of the critical measurement rate and correlation length exponent.} a) Measurement probability $p_c$ determined by crossing of curves, as well as curve collapse, at $N_1$ and $N_2$ sites respectively. The extrapolation is a linear fit to the $p_c$ crossing data. b) Correlation length exponent $\nu$ obtained by fitting two curves of $N_1$ and $N_2$ sites respectively, over $(N_1 N_2)^{-1}$. Confidence intervals are set to the $O = 1.3O^*$ points. All linear fits assign the points equal weight, and omit the rightmost $N_1 = 12, N_2 =16$ point from the fitting. }
    \label{fig:nu_pc_extrapolation}
\end{figure}

We simulate the circuit dynamics for system sizes $N = 12, 16, 20, 24$ over a range of measurement probabilities $p$.
For each pair of sizes, the curves cross at a finite-size estimate $p_c(N_1,N_2)$, as shown in Fig.~\ref{fig:TMI_melko}(a), which we use to locate the critical measurement rate.
We find that the crossing point $p_c(N_1,N_2)$ decreases monotonically with $(N_1N_2)^{-1}$ 
(Fig.~\ref{fig:nu_pc_extrapolation}(a)) - 
with the smallest value at $p_c(20,24)=0.16$, this serves as an upper bound to $p_c$ in the thermodynamic limit, $p_c(\infty) \le 0.16$. 
By linearly extrapolating to $(N_1N_2)^{-1}=0$, we also obtain an estimate $p_c(\infty)=0.142(3)$, where the error analysis can be found in Appendix~\ref{app:TMI_collapse}.

Alternatively, we can obtain both $p_c$ and the 
correlation-length critical exponent $\nu$ by performing a finite-size scaling collapse on the ensemble-averaged $\overline{I_3}$, according to the ansatz~\cite{Pixley2020}
\begin{equation}
    \overline{I_3}(p, N) \sim \mathcal{F}((p-p_c)N^{1/\nu})
\end{equation}
as shown in Fig.~\ref{fig:TMI_melko}(b) for $N=20$ and $24$. The optimal collapse yields $p_c(20,24)=0.160(1)$ and $\nu(20,24)=1.48(4)$.
The optimal $p_c$ for all finite sizes agrees well with the values obtained by TMI crossing, as shown in Fig.~\ref{fig:nu_pc_extrapolation}(a). 
The scaling exponent $\nu$ also scales linearly with $(N_1N_2)^{-1}$, and extrapolates to 
\begin{align} \label{eq:nu}
    \nu(\infty) = 1.34(6)\, . 
\end{align}
as shown in Fig.~\ref{fig:nu_pc_extrapolation}(b).
This exponent is close to the prediction of classical percolation theory $\nu=4/3$~\cite{Skinner2019}.
Compared to previous results on the critical measurement probability and correlation length exponent, Ref.~\cite{Czischek2021}, using the same circuit ensemble, found $p_c = 0.17(2)$ by using the half-chain entanglement entropy, while the correlation-length exponent was estimated to be $\nu=1.4(2)$. For the Haar-random circuit, where the gates are chosen using the Haar measure on $U(4)$, the correlation-length exponent was estimated in Ref.~\cite{Pixley2020} (using the same TMI collapse and $N\leq24$) to be $\nu_{\mathrm{Haar}}=1.2(2)$ at a critical measurement probability of $p_c^{\mathrm{Haar}} = 0.168(5)$.
For the Clifford-random circuit, the best estimate for the exponent is $\nu_{\rm C} = 1.28(2)$~\cite{Gullans2020} - which is still very close to our thermodynamic-limit $\nu$, despite belonging to a different universality class. 

For the subsequent large-scale analysis of spatial correlations, we fix $p=0.17$ and consider scaling with separation. As we demonstrate in the following section, the resulting correlations exhibit robust power-law scaling, consistent with effective criticality in this regime.

\subsection{Scaling Ansatz and Conformal Geometry} \label{sec:scaling_ansatz_and_conformal_geometry}

To extract universal critical exponents that dictate the spatial decay of GME, we analyze the scaling of $k$-party ensemble-averaged  quantities $\overline{E_k}$ (either $\overline{I_k}$ or $\overline{\mathcal{N}_k}$) for a set of disjoint subregions $\{A_1, \dots, A_k\}$. 
However, motivated by the scaling form of $n$-point correlation functions in conformal field theory (CFT)~\cite{francesco2012conformal,Sang2021MIPTNeg}, which exhibit power-law decay governed by the scaling dimensions of the constituent fields, we posit the large-separation ansatz:
\begin{equation} \label{eq:entanglement_exponent_equation}
    \overline{E_k} =  \dist(A_1, \dots, A_k) ^{-\alpha_k}+\cdots \,,
\end{equation}
where $\alpha_k$ is the critical scaling exponent and $\dist$ is the effective conformal distance scale of the subsystem configuration; the ellipsis denote subleading terms. Note that since we utilize single-site subregions ($w=1$), the width dependence is absorbed into the prefactor.
To account for the periodic boundary conditions of the circuit, the Euclidean distance $x$ between two points must be replaced by the conformal chord length $l_x$:
\begin{equation} \label{eq:chord_length_definition}
    l_x = \frac{N}{\pi} \sin \left( \frac{\pi x}{N} \right).
\end{equation}
For two subregions, the unique distance parameter is $\eta^{-1/2}$, where $\eta$ is the conformal cross-ratio of the four points on the ends of the subregions. For multiparty configurations involving $k > 2$ subregions, there is no unique distance parameter, but an effective distance scale $\dist$ can be defined; specifically, we define $\dist$ as the geometric mean of the chord lengths connecting adjacent parties (see Fig.~\ref{fig:circuit}):
\begin{equation} \label{eq:distance_scale}
    \dist(A_1, \dots, A_k) = \prod_{i=1}^k l_{|A_{i+1}-A_i|}^{1/k},
\end{equation}
where $A_{k+1} \equiv A_1$. For $k=2$ this reduces to the distance parameter associated with the conformal cross ratio, $\eta^{-1/2}$.

\section{Results: Spatial Scaling of Multiparty Correlations} \label{sec:results}

\begin{figure}[hbt!]
\centering
\includegraphics[width=0.5\textwidth]{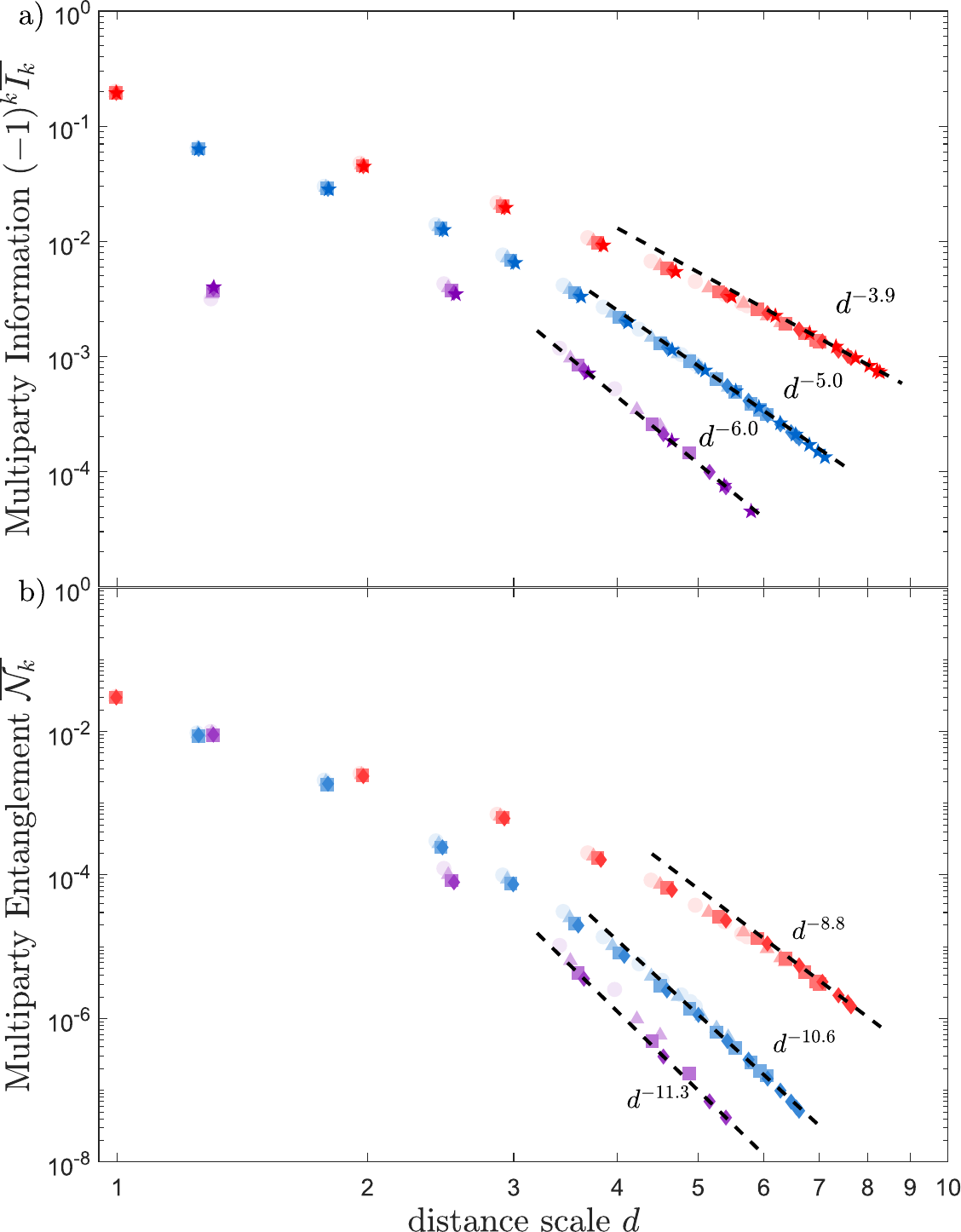}%
\llap{\shortstack{%
    \includegraphics[width=0.17\textwidth]{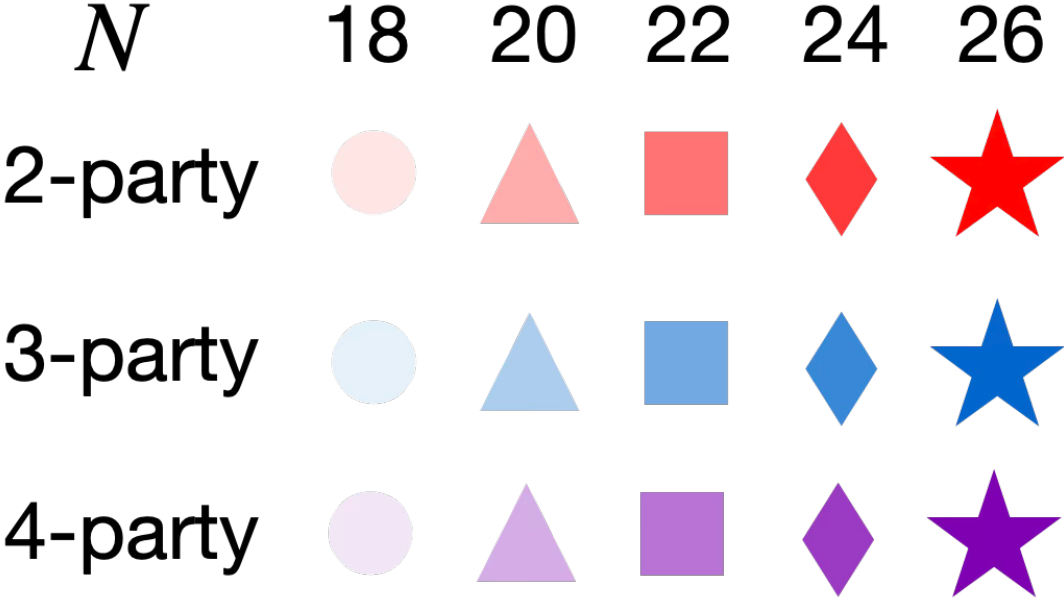}\\
    \rule{0ex}{6.4cm}%
    }
\rule{4.7cm}{0ex}}
\caption{\textbf{Spatial scaling of multiparty correlations.} \textbf{(a)} Multiparty Mutual Information $(-1)^k\overline{I_k}$ and \textbf{(b)} Genuine Multiparty Negativity $\overline{\mathcal{N}_k}$ as a function of the effective distance scale $d$ at $p=0.17$, near the critical point. The number of parties $k$ is distinguished by symbol and color: 2-party (red circles), 3-party (blue triangles), and 4-party (purple squares). Color intensity indicates system size, scaling from light ($N=18$) to dark ($N=26$). Dashed black lines denote power-law fits extracted from the largest converged dataset for each metric: $N=26$ for MI and $N=24$ for GMN. }
\label{fig:ENT_dist_scaling}
\end{figure}

Having established the critical nature of the system, we now investigate the spatial architecture of multiparty correlations at $p=0.17$, near criticality. Figure~\ref{fig:ENT_dist_scaling} displays the spatial scaling of the Multiparty Mutual Information and the Genuine Multiparty Negativity as a function of the effective distance scale $d$.
We analyze the spatial scaling by measuring correlations among qubits separated by a distance $x$. Specifically, we consider symmetric clusters of $k=2,3,4$ equidistant qubits, whose indices are $\{i, i+x\}$, $\{i, i+x, i+2x\}$, and $\{i, i+x, i+2x, i+3x\}$ respectively. We also consider an asymmetric 3-qubit configuration featuring unequal spacings of $x$ and $x+1$, with indices $\{i, i+x, i+2x+1\}$.

In our characterization of the Multiparty Mutual Information, we find that it follows an alternating sign hierarchy: $(-1)^k \overline{I_k} > 0$ for $k=2,3,4$. In accordance with this hierarchy, the tripartite mutual information is consistently negative for all system sizes and distances studied; in Fig.~\ref{fig:ENT_dist_scaling}(a), we therefore plot the negative tripartite mutual information $-\overline{I_3}$ and the four-party mutual information $\overline{I_4}$.

\begin{figure}[hbt!]
    \centering
    \begin{tikzpicture}
        \begin{scope}
            \node[anchor=north west,inner sep=0] (image_a) at (0,3)
            {\includegraphics[width=0.95\columnwidth]{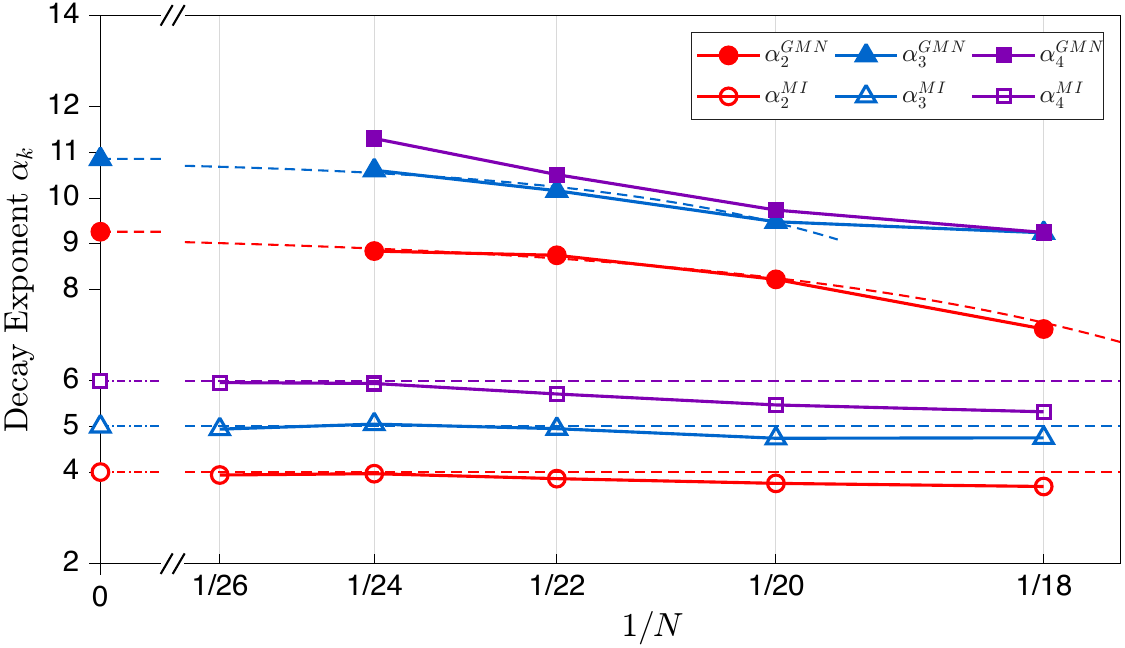}};
        \end{scope}
    \end{tikzpicture}
    \caption{\textbf{Large-$N$ convergence of multiparty exponents.} Finite-size estimates of the decay exponents for genuine multiparty negativity ($\alpha_k^{\mathrm{GMN}}$) and multiparty mutual information ($\alpha_k^{\mathrm{MI}}$) plotted versus $1/N$, using the fitted values in Table.~\ref{tab:entanglement_exponents}. For mutual information, the exponents show negligible drift up to $N=26$, and we therefore plot the expected asymptotic constants $\alpha_k^{\mathrm{MI}}=k+2$. For GMN, the $k=2,3$ exponents retain visible finite-size drift and are extrapolated with $\alpha_k^{\mathrm{GMN}}(N)=C-A e^{b/N}$ (dashed lines) for $N\geq 20$. For $\alpha_4^{\mathrm{GMN}}$, the available system sizes are insufficient to support a reliable extrapolation.}
    \label{fig:exponents_over_N}
\end{figure}

Crucially, our data reveals a robust algebraic decay for both mutual information and negativity across all $k=2,3,4$ parties. This establishes long-range multiparty entanglement as a characteristic feature of this dynamical critical phase. 
To quantify the distinct scaling behaviors, we extract the decay exponents $\alpha_k^{\rm{MI}}, \alpha_k^{\rm{GMN}}$ by fitting the large distance data to the power-law form (\ref{eq:entanglement_exponent_equation}) using weighted least squares with a minimum relative error of 0.03. The evolution of these exponents as a function of system size $N$ is plotted in Fig.~\ref{fig:exponents_over_N}, with specific values summarized in Table~\ref{tab:entanglement_exponents}. The fitting ranges used for each system size and number of parties are listed in Appendix Table~\ref{tab:fit_ranges}. 
Finally, we note that the fitted exponents increase monotonically with the system size. This trend is expected in a periodic chain: at fixed separation, increasing $N$ dilutes correlations across a larger number of qubits, decreasing the entanglement between specific subregions. As a result, the largest-$N$ values in Table~\ref{tab:entanglement_exponents} already provide conservative lower bounds on the thermodynamic exponents.

Our fitted exponents are broadly consistent with, but slightly larger than, values reported previously. One recent study reports $\alpha_2^{\mathrm{GMN}}=7.1$, and $\alpha_3=8.7-8.8$ for alternative tripartite entanglement quantifiers $\mathcal I_2$ and $W_3$ (which are not GME monotones)~\cite{Sebastien2025}. An earlier study~\cite{Sang2021MIPTNeg} also found a smaller bipartite exponent $\alpha_2^{\mathrm{GMN}}=6.2$.
We attribute the differences primarily to the numerical difficulty of extracting asymptotic exponents from a tail-dominated observable; see Appendix~\ref{app:decomposition_entanglement} for the dominant algebraic decay of the entangling probability.
Our large-scale simulations are designed to overcome the slow convergence of tail-dominated GMN scaling by providing sufficient system size and statistics to isolate a stable asymptotic window, and we therefore expect our fitted exponents to be among the most accurate available so far. 
Moreover, GMN is evaluated via a semidefinite program, yielding a globally optimal entanglement monotone with guaranteed convergence (unlike $\mathcal I_2$ and $W_3$, which are not entanglement monotones and rely on nonlinear optimization). 
In terms of mutual information, the scaling of the 2-party MI, $\alpha_2^{\mathrm{MI}}=4$, agrees with previous studies of the critical Haar-random circuit~\cite{Skinner2019,Li2019_MIPT,Shi2020,Sang2021MIPTNeg}. 
It would be interesting to understand whether the minimal-cut picture in Ref.~\cite{Skinner2019} can be extended to account for the higher-party trend $\alpha_k^{\mathrm{MI}}\approx k+2$.

After extracting the spatial decay exponents $\alpha_k$, we next investigate how these exponents evolve with the number of parties $k$. Fig.~\ref{fig:exponents_over_N} highlights an important distinction between total correlations and entanglement in this phase. The multiparty mutual information exponents converge rapidly with $N$, stabilizing by $N=26$ near simple integer values close to 
\begin{align}
    \alpha_k^{\rm{MI}}=k+2
\end{align}
In contrast, the GMN exponents exhibit a stronger drift with system size and tend to increase with $N$, suggesting that the asymptotic power-law regime for multiparty entanglement requires longer distances and larger systems to fully manifest. This slower convergence is consistent with the markedly different statistics of the two observables: while multiparty mutual information is governed by typical correlations, the negativity is dominated by rare long-range entanglement events and exhibits a heavy-tailed distribution. Consequently, our GMN scaling analysis focuses on the largest converged data set $(N=24)$; realization counts are listed in Appendix Table.~\ref{tab:rdm_circuit_counts}.

\subsection{Comparison to Theoretical Constraints}\label{sec:exp_lower_bound}

A defining feature shared by both the mutual information and the negativity is their algebraic decay with distance, which is in drastic contrast with typical equilibrium systems, where entanglement is extremely short-ranged~\cite{Osterloh2002,Osborne2002,Javanmard2018,parez2024fate}. In the one-dimensional transverse field Ising model, entanglement vanishes exactly beyond next-nearest neighbors. In two and three dimensions, entanglement disappears as soon as spins are not adjacent~\cite{Wang2025EntMicro,Lyu2025Ising}. 
This pronounced locality of entanglement is commonly attributed to quantum monogamy~\cite{CKW2000}, which constrains the extent to which a quantum system can simultaneously share entanglement with multiple partners. 
For instance, a pure state consisting of $N+1$ qubits satisfies the CKW inequality $N_{A|B_1}^2 + ... +N_{A|B_N}^2\leq N_{A|B_1...B_N}^2\leq1/4$~\cite{CKW_negativity2007} which bounds the total bipartite entanglement that subsystem $A$ can distribute among the remaining qubits.
In typical equilibrium systems, all forms of multiparty entanglement remain short-ranged~\cite{parez2024fate}, and such monogamy inequalities are comfortably satisfied.
The algebraic decay observed in the measurement-induced critical phase therefore reflects a fundamentally different organization of multiparty correlations, one that emerges in a non-equilibrium dynamical setting and remains compatible with monogamy constraints. In particular, for bipartite negativity between qubits, quantum monogamy imposes a lower bound on the decay exponent, 
$\alpha_2^{\mathrm{GMN}}>1/2$. 
This lower bound is easily satisfied by the scaling observed in Fig.~\ref{fig:ENT_dist_scaling}. 

We next turn to the monogamy of multipartite mutual information, which yields a particularly restrictive constraint on long-range total correlations. 
In our simulations, we observe a clear alternating-sign hierarchy of the ensemble-averaged multipartite mutual information across different scales.
For microscopic subregions in which each party consists of a single qubit, the tripartite mutual information is negative $(\overline{I_3}<0)$ and the four-partite mutual information is positive $(\overline{I_4}>0)$ across all system sizes and distances studied.
In addition, we also observe negative tripartite mutual information for macroscopic subregions, consistent with earlier findings~\cite{Pixley2020}.
Taken together, these microscopic and macroscopic observations motivate the working assumption that the multipartite mutual information is sign definite for distant, microscopic subregions.
We note in passing that alternating-sign multipartite mutual information appears in holographic systems~\cite{Hayden2013_multipartyMI,Mirabi2016_holography_MI_monogamy}, and that holographic connections to measurement-induced criticality have been discussed previously~\cite{Li2019_MIPT,Bao2020_MIPT}.
Under this assumption, and for configurations consisting of $k$ microscopic subregions of linear size $\ell$ separated by distances $r \gg \ell$, monogamy imposes a quantitative constraint on spatial scaling. Assuming scale invariance at criticality, algebraic decay of the form $\overline{I_k}(r) \sim 1/r^{\alpha_k^{\rm{MI}}}$ cannot be arbitrarily slow, leading to a lower bound 
\begin{align} \label{eq:boundMI}
    \alpha_k^{\rm{MI}} \geq k\, . 
\end{align}
As we find $\alpha_k^{\rm{MI}} \approx k+2$ empirically for $k=2,3,4$ this constraint is comfortably satisfied. Derivations of both the multipartite mutual information bound and the bipartite negativity bound are provided in Appendix~\ref{app:monogamy_lower_bound}.

Several constraints have been conjectured for the scaling of bipartite and multipartite entanglement, motivated by earlier studies of critical non-unitary systems.
A recent work~\cite{allen2025spatial} proposed a set of constraints for genuine multiparty entanglement in monitored critical systems: the exponents should (i) be larger than those of total correlations, $\alpha_k \ge \alpha_k^{\mathrm{MI}}$; (ii) increase monotonically with $k$, $\alpha_{k+1}\geq\alpha_k$; and (iii) satisfy a subadditivity condition, $\alpha_{k+\ell} \le \alpha_k + \alpha_\ell$. 
For systems that obey (\ref{eq:boundMI}), (i) implies:
\begin{align}
    \alpha_k \geq k
\end{align}

In the special case of a measurement-only circuit describing a projective Ising model, the subadditivity bound (iii) is saturated, yielding an exactly additive spectrum $\alpha_k =2 k$~\cite{allen2025spatial}. 
Our scaling exponents in Table.~\ref{tab:entanglement_exponents} satisfy all three conjectures for every system size $N$. 
In particular, we observe \textit{strict} subadditivity: $\alpha_4 < 2\alpha_2$ for all values of $N$ measured. Additionally, we observe $\alpha_3 < \frac{3}{2}\alpha_2$ not only for all values of $N$ measured, but for the extrapolated large-$N$ exponents as well.

\subsection{Entanglement-weighted graphs}\label{sec:ent_weight_graphs}

\begin{figure}[hbt!]
    \begin{tikzpicture}
        \begin{scope}
            \node[anchor=north west,inner sep=0] (image_a) at (0,0)
            {\includegraphics[width=\columnwidth]{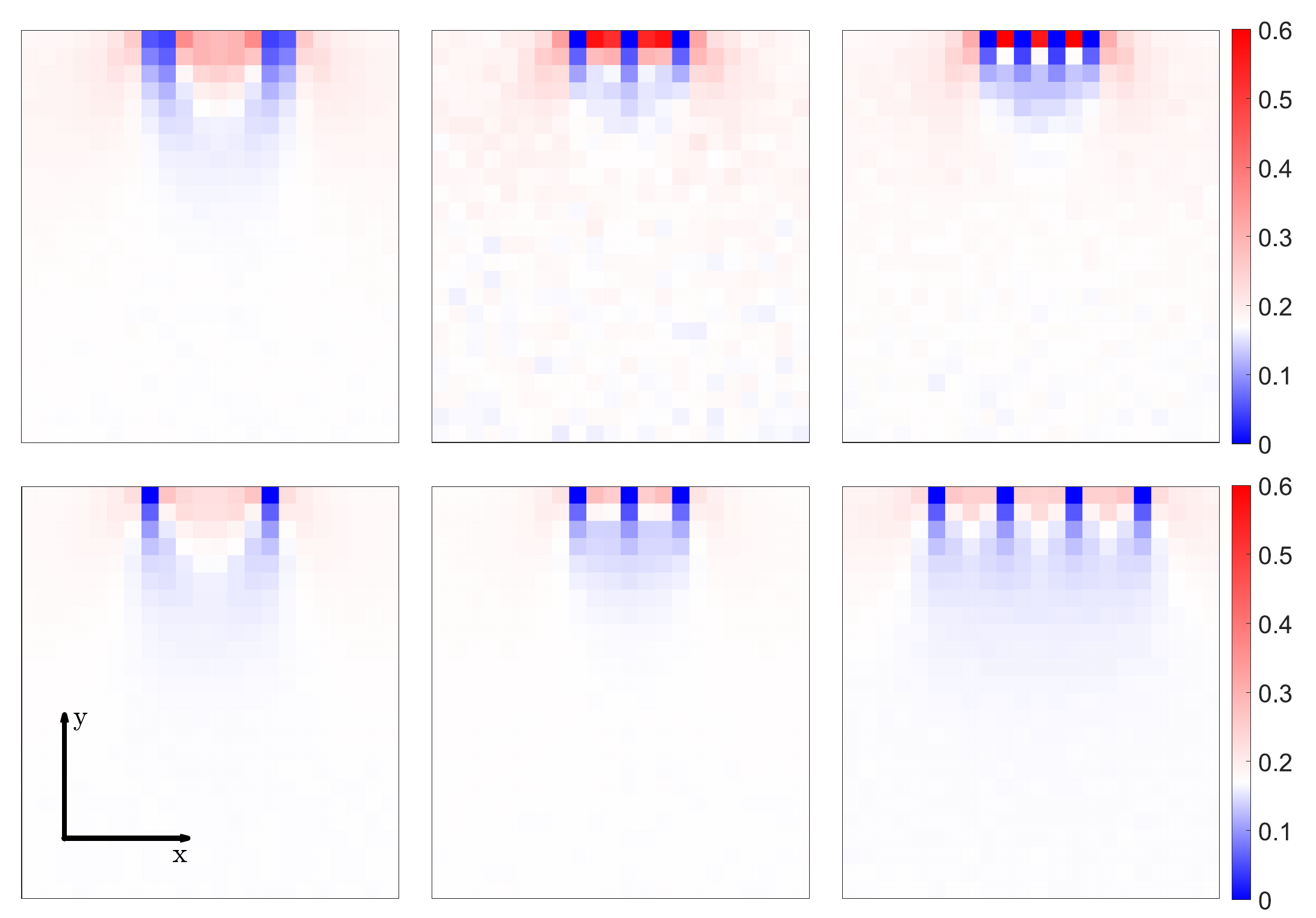}};
            \node [anchor=north west] (note) at (0.1,-2.45) {\small{\textbf{a)}}};
            \node [anchor=north west] (note) at (2.8,-2.45) {\small{\textbf{b)}}};
            \node [anchor=north west] (note) at (5.5,-2.45) {\small{\textbf{c)}}};
            \node [anchor=north west] (note) at (0.1,-5.45) {\small{\textbf{d)}}};
            \node [anchor=north west] (note) at (2.8,-5.45) {\small{\textbf{e)}}};
            \node [anchor=north west] (note) at (5.5,-5.45) {\small{\textbf{f)}}};
        \end{scope}
    \end{tikzpicture}
    \centering
    
    \caption{\textbf{Entanglement- and MI-weighted graphs.} Top row (\textit{a}-\textit{c}) displays the Genuine Multiparty Negativity at distances $x=6, 3, 2$ respectively. Bottom row (\textit{d}-\textit{f}) displays the Multiparty Mutual Information at distances $x=7, 4, 4$. All data corresponds to a system size $N=22$ at $p=0.17$. The color gradient indicates the entanglement weighted local measurement rate, where white corresponds to a baseline value $0.17$, while blue (red) represents lower (higher) rates; the disjoint blue regions at the top layer correspond to the distinct parties. Note that panel (\textit{a}) utilizes 2-site subregions per party, while all other configurations use single-site parties.}
    \label{fig:ewg_stacked_neg_mi}
\end{figure}
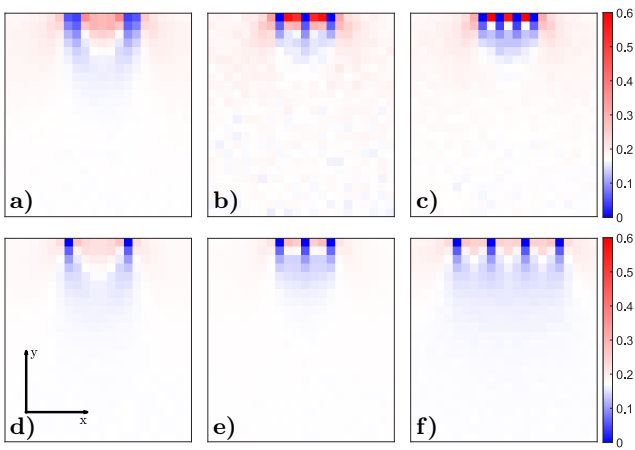

To understand the geometric origins of the rare multiparty entanglement events identified in the previous section, we must look at the conditions that would produce a genuinely multiparty entangled system to begin with. In a random circuit, the creation of an entangled state is driven by spacetime heterogeneities—specific fluctuations in the measurement record that create protected paths for quantum information. To visualize these fluctuations, we utilize Entanglement-Weighted Graphs (EWG)~\cite{allen2025spatial}, which reveal the average spacetime structures supporting multiparty correlations. 
Formally, the EWG represents the weighted-average structure of the circuit ensemble, where the contribution of each realization is weighted by its generated $k$-party entanglement (or mutual information). We start with a lattice variable $W_\varepsilon(x,t)$ that represents the structure of a specific circuit configuration $\varepsilon$. In our case, $W_\varepsilon(x,t)$ represents the binary measurement record of the configuration, with $W_\varepsilon(x,t)=1$ denoting a measurement at a specific spacetime point and $0$ denoting the absence of a measurement. The weighted graph is then given by:
\begin{gather}
    \overline{W}[E_k](A_1, ..., A_k) = \frac{\sum_\varepsilon E_k(A_1, ... ,A_k | \varepsilon) \cdot W_\varepsilon}{\sum_\varepsilon E_k(A_1, ... ,A_k | \varepsilon)}
\end{gather}

Figure~\ref{fig:ewg_stacked_neg_mi} presents the resulting weighted graphs for a system of size $N=22$ at the critical point $p=0.17$. 
The color gradient depicts the local measurement rate relative to this baseline: white regions correspond to $p=0.17$, while blue and red indicate lower and higher rates, respectively. 
The top row (panels \textit{a--c}) illustrates the geometry of Genuine Multiparty Negativity, which requires a strikingly distinct structure. For GMN, the parties, visible as disjoint blue regions at the top boundary, are connected by a pristine, low-measurement ``bulk'' (blue). Crucially, this bulk is surrounded by a ``halo'' of high measurement density (red). This halo effectively isolates the entangled cluster from the rest of the system, preventing the monogamous quantum correlations from leaking into the environment. The necessity of these specific geometric structures explains the rarity of GMN events noted earlier: resolving these features requires filtering millions of realizations to find the rare few that satisfy these constraints. 

The MI-weighted graphs (Fig.~\ref{fig:ewg_stacked_neg_mi} \textit{d--f}) display an overall spatial organization that closely mirrors the GMN-weighted graphs, suggesting that information sharing is also structured in a monogamy-like manner in the hybrid critical phase. This behavior is qualitatively different from measurement-only circuits~\cite{allen2025spatial}, where mutual information shows no clear positive correlation with the local measurement pattern. The emergence of a correlated MI--measurement structure therefore provides a direct visual signature of the nontrivial role played by unitary dynamics in hybrid circuits.
Despite this qualitative similarity, the ``low-measurement'' halo is noticeably more diffuse in MI than in GMN. Statistically, this reflect more restrictive conditions to sustain a finite GMN:  realizations that yield a finite GMN are much rarer than for the MI.
A quantitative comparison of the distribution of GMN and MI for a specific configuration is shown in Appendix Fig.~\ref{fig:nonzero_histogram}, while a comprehensive list of nonzero-event counts across system sizes can be found in Appendix Table.~\ref{tab:count_count}.

\section{Conclusion} \label{sec:conclusion}
We have studied the GME of disjoint parties near an MIPT belonging to the Haar non-unitary CFT class. We showed that the gate set of our circuit ensemble, beyond its relative simplicity and experimental connections to trapped-ion architectures, also shows better ensemble convergence compared to the continuous Haar random case often studied. 
The GME exponents up to $k=4$ parties were estimated via finite-size scaling, and were shown to obey non-trivial inequalities. We also studied the multiparty mutual information, and our robust results lead to a conjecture for the critical exponents $\alpha_k^{\rm MI} = k+2$, which obeys the monogamy lower bound we put forth: $\alpha_k^{\rm MI}\geq k$. 

Going forward, it would be interesting to better converge the 4-party GME exponents. Although we have focused on equal separations in that case, one could consider less-symmetric configurations to get more points on the curve.
Furthermore, such analysis could be performed at distinct MIPTs to better understand what determines the entanglement structure. 
A crucial question comes to mind: could one predict the GME and MI exponents from the underlying CFT?

\section*{Acknowledgements}
We are grateful to S.~Avakian and T.~Pereg-Barnea for useful discussions. We also wish to thank the staff at Calcul Qu\'ebec, Universit\'e de Montr\'eal, and Perimeter Institute for their technical assistance.
The research has been funded by a NSERC Quantum Alliance grant.
W.W.-K.\/ and L.L.\/ are supported by a grant from the Fondation Courtois, a Chair of the Institut Courtois, a Discovery Grant from NSERC, and a Canada Research Chair. Computational resources were provided by the Digital Research Alliance of Canada (Beluga, Cedar, Narval, Nibi, Rorqual, and Fir) and by Perimeter Institute through access to the Symmetry cluster.
Research at Perimeter Institute is supported in part by the Government of Canada through the Department of Innovation, Science and Economic Development Canada and by the Province of Ontario through the Ministry of Economic Development, Job Creation and Trade.

\bibliographystyle{longapsrev4-2}
\bibliography{bibtex}

\clearpage
\onecolumngrid

\begin{appendices}

\textbf{\Large Supplemental Materials for Taming multiparty Entanglement at measurement induced phase transitions}

\renewcommand{\thesubsection}{\arabic{subsection}}

\section{TMI Scaling Collapse Analysis for $p_c$ and $\nu$}\label{app:TMI_collapse}

\begin{figure}[hbt!]
    \centering
    \includegraphics[width=0.95\columnwidth]{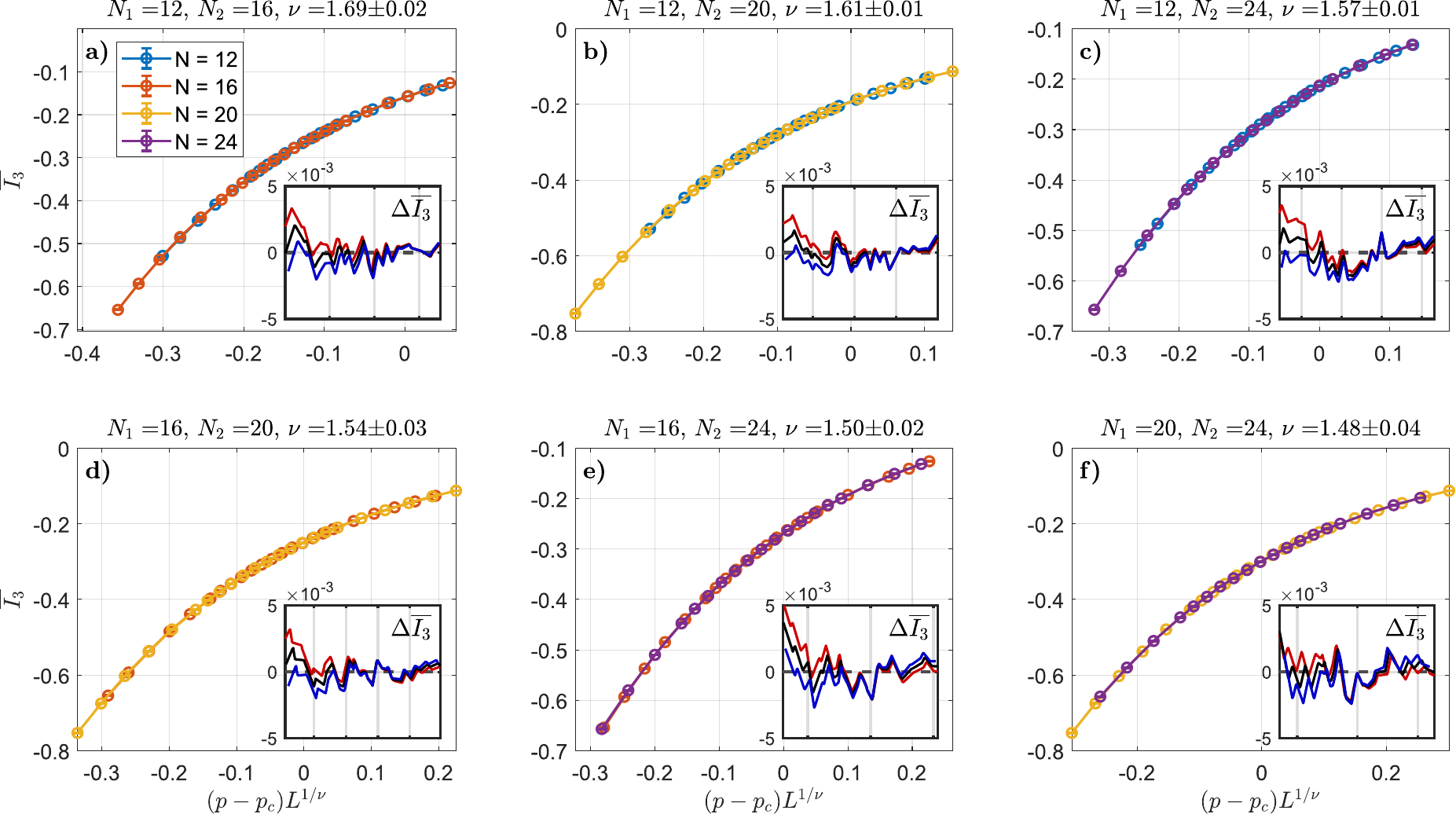}
    \caption{\textbf{Curve collapse of all pairs of $N$ values.} Each curve is a plot of ensemble-averaged TMI over the rescaled distance from the critical point $(p-p_c^*)L^{1/\nu^*}$, at the optimum $p_c^*, \nu^*$ that minimizes the objective function at $O^*$. \textit{Inset:} Collapse-diagnostic based on the difference $\Delta \overline{I_3}$ between the two curves, at $\nu^*-\Delta\nu, \nu^*, \nu^*+\Delta\nu$ (red, black and blue respectively).}
    \label{fig:TMI_curve_collapse_N1N2}
\end{figure}

\begin{table}[h]
\centering
\begin{tabular}{l|ccccccc}
\hline
$\mathbf{(N_1, N_2)}$ & \textbf{(12,16)} & \textbf{(12,20)} & \textbf{(12,24)} & \textbf{(16,20)} & \textbf{(16,24)} & \textbf{(20,24)}\\
\hline
$O^*$& 1.04 & 1.09 & 2.44 & 0.83 & 1.81 & 2.08\\
\hline
$p_c^*$, cross  & 0.189 & 0.179 & 0.172 & 0.167 & 0.165 & 0.160 \\ 
\hline
$p_c^*$, collapse & 0.189 & 0.178 & 0.173 & 0.168 & 0.164 & 0.160 \\
\hline
$\nu^*$& 1.69 & 1.61 & 1.57 & 1.54 & 1.50 & 1.48 \\
\hline
\end{tabular}
\caption{TMI curve collapse analysis for each pair of $N$ values. $O^*$ denotes the objective function minimum $O^*$, which should be close to 1~\cite{Pixley2020}. The collapse measurement probability $p_c^*$ and scaling exponent $\nu^*$ minimize the objective function for each pair. The $p_c^*$ obtained from curve crossing is also recorded.}
\label{tab:curve_collapse_parameters}
\end{table}

To obtain the effective critical measurement probability $p_c^*(N_1, N_2)$ for each pair of system sizes $N_1, N_2$, we can either find the point where their curves cross, or we can perform a curve collapse to find the optimal values $p_c^*, \nu^*$ that minimize the objective function $O(p_c, \nu)$ from Ref.~\cite{Pixley2020}.
Using the crossing approach, the critical measurement rate $p_c^*(N_1, N_2)$ is determined by the intersection of curves for each pair of system sizes, calculated via both linear and spline interpolations. We report the value derived from the spline fit, while the absolute difference between the two interpolation methods provides an estimate of the systematic error. This is then summed with a statistical error component obtained from the four data points surrounding the crossing. This statistical uncertainty is estimated by assuming maximal correlation, wherein the surrounding points are shifted by their standard errors in coordinated, opposing directions to maximize the variation in $p_c^*$, providing a conservative bound on the total uncertainty.
Using the collapse approach, the error in $p_c^*$ is determined by the points where the objective function $O(p_c, \nu^*) = 1.3O^*$ ($O^*$ is the minimum value), as per Ref.~\cite{Pixley2020}. The correlation length exponent $\nu^*(N_1, N_2)$ for each curve is determined solely by minimizing the objective function, with its errors set by $O(p_c^*, \nu) = 1.3O^*$ as well.

Due to the large discrepancy between the errors in individual optimized values of $p_c^*(N_1, N_2)$, $\nu^*(N_1, N_2)$ and their residuals from the linear fit, we used a bootstrap-inspired procedure to estimate errors in the infinite-$N$ extrapolated values of $\nu$ and $p_c$. For each pair $(N,p)$ simulated, a sample of 60,000 circuits (less than half of the total sample size at each point - see Table~\ref{tab:TMI_circuit_count} of Appendix~\ref{app:sample_sizes} for the precise sample sizes) is randomly drawn with replacement, creating a reduced data set. From this data set, the infinite-$N$ extrapolated values of $\nu$ and $p_c$ are measured. We then repeat this process for 200 sets of extrapolated values and measure their standard deviation: 0.06 for $\nu$ and $0.001-0.003$ for the curve-collapse and curve-crossing $p_c$, respectively. 
Because these standard deviations are derived from smaller bootstrapped subsamples rather than the full data set, they provide a conservative upper bound for the error of the extrapolated values.

\section{Decomposition of Average Entanglement into Probability and Magnitude} \label{app:decomposition_entanglement}

\begin{table}[h]
\centering
\begin{tabular}{l|cccccc}
\hline
\textbf{N} & \textbf{18} & \textbf{20} & \textbf{22} & \textbf{24} & \textbf{26}\\
\hline
Circuits (millions)  & 6.81 & 10.71 & 15.59 & 26.28 & 0.166 \\ 
\hline
RDMs (millions)   & 122.7 & 214.2 & 343.0 & 589.2 & 4.316 \\
\hline
\end{tabular}
\caption{Circuits and effective number of reduced density matrices per non-maximal separation, for each value of $N$. Convergence is obtained for negativity and mutual information exponents for $N \leq 24$, and for mutual information exponents only at $N = 26$.}
\label{tab:rdm_circuit_counts}
\end{table}

\begin{figure}[hbt!]
    \centering
    \includegraphics[width=0.49\textwidth]{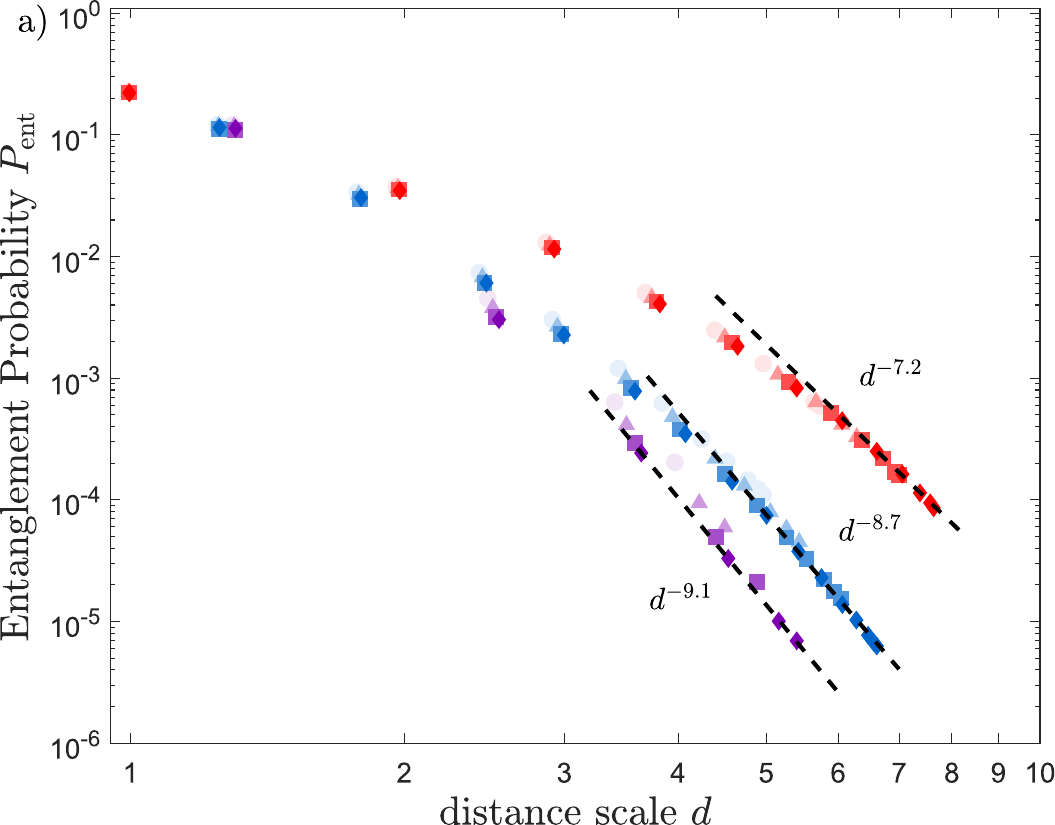}
    \llap{\shortstack{
        \includegraphics[width=0.13\textwidth]{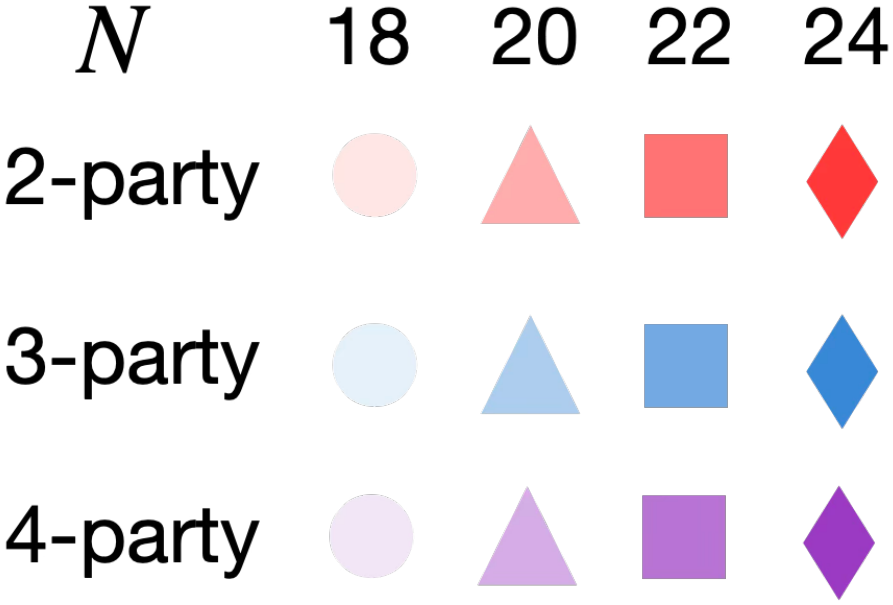}\\
        \rule{0ex}{1.2cm}
        }
    \rule{5.0cm}{0ex}} 
    \hfill
    \includegraphics[width=0.49\textwidth]{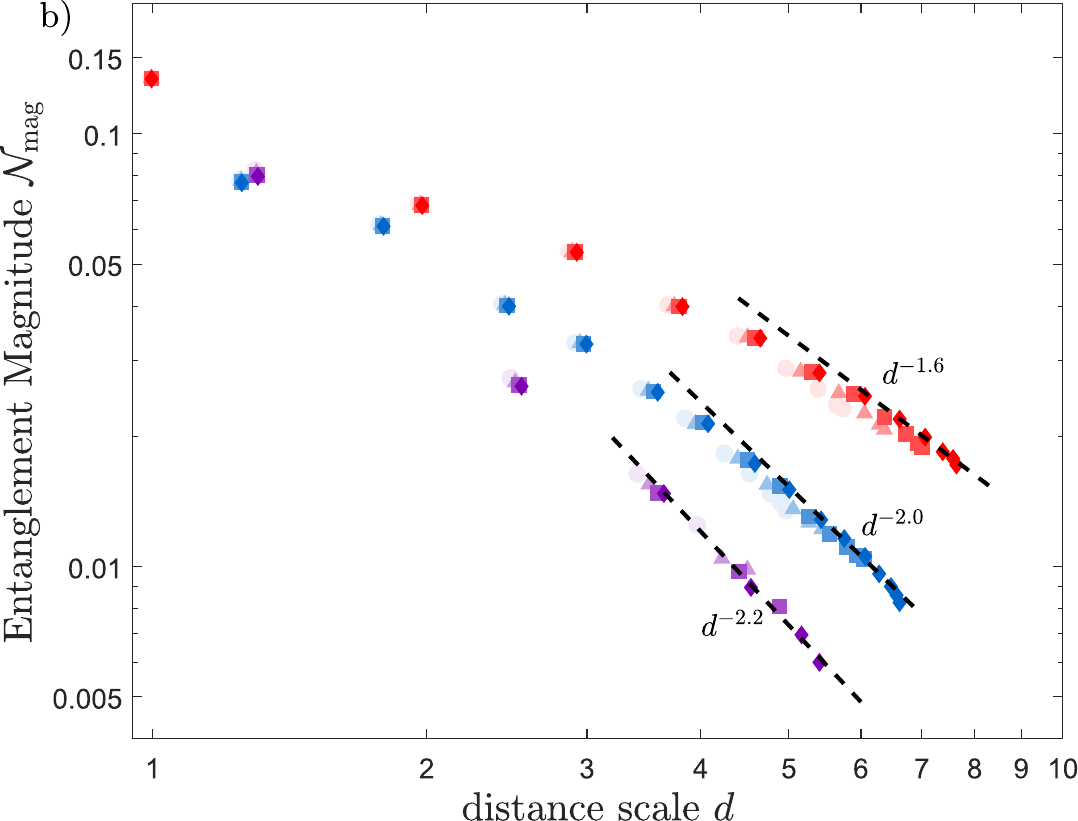}
    
\caption{\textbf{Decomposition of entanglement scaling.} 
    \textbf{(a)} The entanglement probability $P_{\mathrm{ent}}$ and 
    \textbf{(b)} the average entanglement magnitude $\mathcal{N}_{\mathrm{mag}}$ (conditioned on $\mathcal{N}>0$) as a function of the distance scale $d$. 
    Data is shown for various system sizes (indicated by opacity). 
    The dashed lines indicate power-law fits performed specifically on the $N=24$ dataset.
    }
    \label{fig:scaling_prob_mag}
\end{figure}

To understand the physical origin of the observed GME scaling, we introduce the \emph{entanglement magnitude}, $\mathcal{N}_{\mathrm{mag}}$, defined as the average Genuine Multiparty Negativity conditioned on the state being entangled (i.e., $\mathcal{N} > 0$). The total average negativity $\overline{\mathcal{N}_k}$ shown in Fig.~\ref{fig:ENT_dist_scaling} and discussed in the main text can thus be decomposed into the product of the probability of finding an entangled state, $P_{\mathrm{ent}}$, and this entanglement magnitude: 
\begin{equation}
    \overline{\mathcal{N}_k}(d) = P_{\mathrm{ent}}(d) \times \mathcal{N}_{\mathrm{mag}}(d)
\end{equation}

The scaling of these two components is presented in Fig.~\ref{fig:scaling_prob_mag}. We find that the rapid decay of the total entanglement is primarily driven by the sparsity of entangled states rather than a vanishing magnitude. Specifically, the probability $P_{\mathrm{ent}}$ decays with a steep exponent (e.g., $\gamma_p \approx 7.2$ for $k=2$), whereas the entanglement magnitude $\mathcal{N}_{\mathrm{mag}}$ decays significantly slower ($\gamma_{\mathrm{mag}} \approx 1.6$). This decomposition also motivates the specific fitting ranges used in our analysis. While the probability $P_{\mathrm{ent}}$ [Fig.~\ref{fig:scaling_prob_mag}(a)] follows a robust power law up to the largest distances, the entanglement magnitude for the 3-party case ($N=24$) in Fig.~\ref{fig:scaling_prob_mag}(b) exhibits a downward deviation around $d \approx 6.4$. Since $P_{\mathrm{ent}}$ remains smooth in this region, we attribute this downturn to finite-sampling effects where entangled events become too rare to fully converge the average magnitude. 

\begin{figure}[hbt!]
    \centering
    \begin{tikzpicture}
        \begin{scope}
            \node[anchor=north west,inner sep=0] (image_a) at (0,0)
            {\includegraphics[width=0.9\columnwidth]{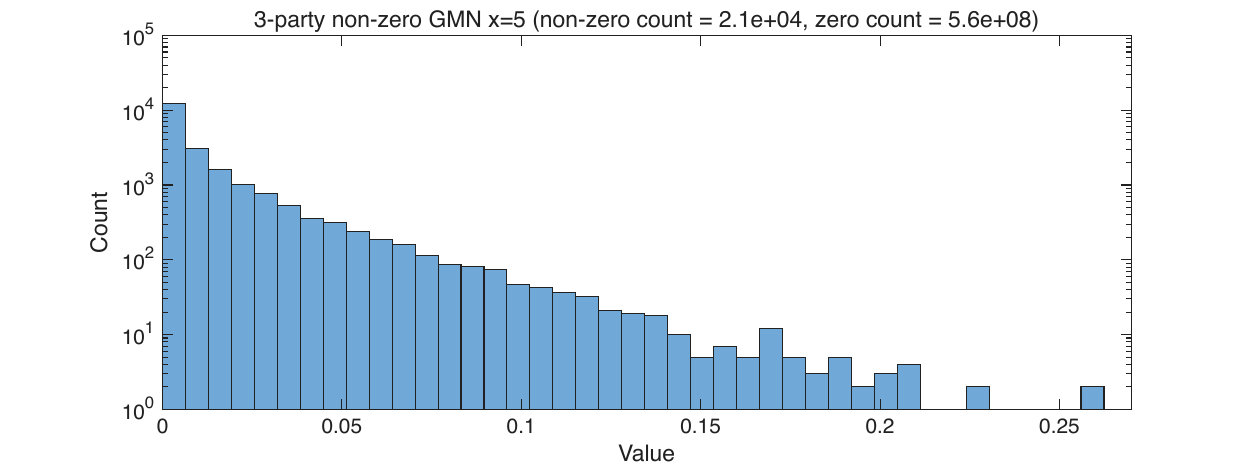}};
            \node [anchor=north west] (note) at (0.8,-0.2) {\small{\textbf{a)}}};
        \end{scope}
    \end{tikzpicture}
    \begin{tikzpicture}
        \begin{scope}
            \node[anchor=north west,inner sep=0] (image_a) at (0,0)
            {\includegraphics[width=0.9\columnwidth]{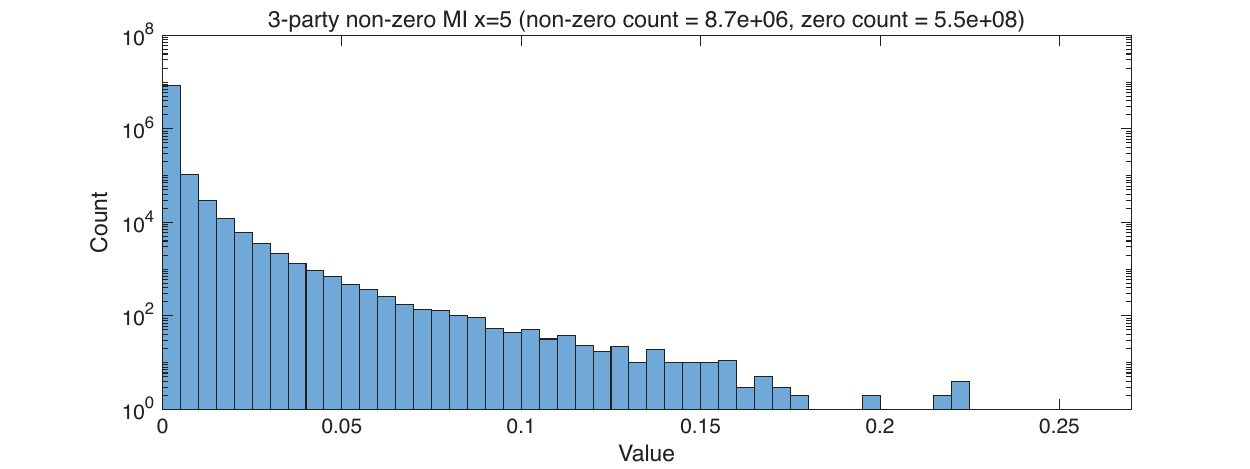}};
            \node [anchor=north west] (note) at (0.8,-0.2) {\small{\textbf{b)}}};
        \end{scope}
    \end{tikzpicture}
   \caption{\textbf{Distribution of nonzero entanglement and mutual information.} Histograms of (a) 3-party genuine multiparty negativity (GMN) and (b) tripartite mutual information (TMI) for subregions at separation $x=5$. Both distributions are highly skewed, with the frequency of large values decaying rapidly; counts are shown on a logarithmic scale to resolve the heavy-tailed behavior. These histograms are generated from a representative subset of the realizations at $N=24$. From the same total sample size, the GMN yields significantly fewer nonzero instances than the TMI, demonstrating the extreme statistical sparsity of genuine multiparty entanglement compared to total correlations at large separations.}
    \label{fig:nonzero_histogram}
\end{figure}

To gain further insight into the statistical nature of these observables, we examine the distribution of realizations at a fixed qubit configuration. Figure~\ref{fig:nonzero_histogram} provides a representative ``zoom-in'' of the tripartite entanglement and mutual information for $N=24$ at a distance of $x=5$. For the same total sample size, the GMN yields significantly fewer nonzero hits than the TMI, quantifying the extreme sparsity that drives the conditional magnitude $\mathcal{N}_{\mathrm{mag}}$. Furthermore, both distributions are highly skewed; as shown in Fig.~\ref{fig:nonzero_histogram}, the nonzero values are concentrated at small magnitudes with a heavy tail of rare, high-entanglement events. 

\section{Fitting ranges for the decay exponents}~\label{app:fitting_ranges}

\begin{table}[h]
\centering
\begin{tabular}{r|c|c|c}
\hline
$\,$\textbf{Entanglement type} $\,$ & $\,$ 2-party $\,$ & $\,$ 3-party $\,$ & $\,$ 4-party $\,$\\
\hline
Lower fit range $\,$ & 6.2 & 4.7 & 2.8 \\ 
\hline
Upper fit range $\,$ & 8.2 & 6.5 & 5.5 \\
\hline
\end{tabular}
\caption{Fitting ranges for the decay exponents in Table.~\ref{tab:entanglement_exponents}. }
\label{tab:fit_ranges}
\end{table}

Table~\ref{tab:fit_ranges} summarizes the standard coordinate intervals used to extract the decay exponents $\alpha_k$ presented in Table~\ref{tab:entanglement_exponents}. The selection of these windows is dictated by the requirement to balance two competing numerical factors. First, the distance scale $d$ must be sufficiently large to reach the asymptotic regime described by the power-law ansatz in Eq.~\ref{eq:entanglement_exponent_equation}. Second, the statistical sparsity of entanglement events at large separations introduces significant noise. Consequently, the fitting windows are restricted to regions where the sample size remains sufficient to yield a robust average value. While most fits adhere to these ranges, specific 2-party configurations require the following adjustments:

\begin{itemize}
\item \textbf{2-party mutual information:} For $N = 18$ and $20$, exponents are extracted using the four points at the largest distances. For $N \ge 22$, the ranges in Table~\ref{tab:fit_ranges} are applied.
\item \textbf{2-party negativity:} Fits consistently incorporate the four largest distance scales across all system sizes. At $N=24$, the largest distance point is excluded due to insufficient sample size.
\end{itemize}

\section{Comparison with Haar Circuit}\label{app:compare_haar}

In this Appendix we provide a benchmark of our MMS-circuit data against a generic Haar-random hybrid circuit at the same finite system size, $N=18$. At this size the GMN data is far from its asymptotic regime, so the comparison is not intended to establish thermodynamic-limit exponents. Rather, it serves two modest purposes: (i) to highlight qualitative similarities and differences between the two ensembles at fixed $N$, and (ii) to motivate our choice of the MMS circuit as the primary platform for the exponent analysis in the main text.

\begin{figure}[hbt!]
    \centering
    \includegraphics[width=0.5\textwidth]{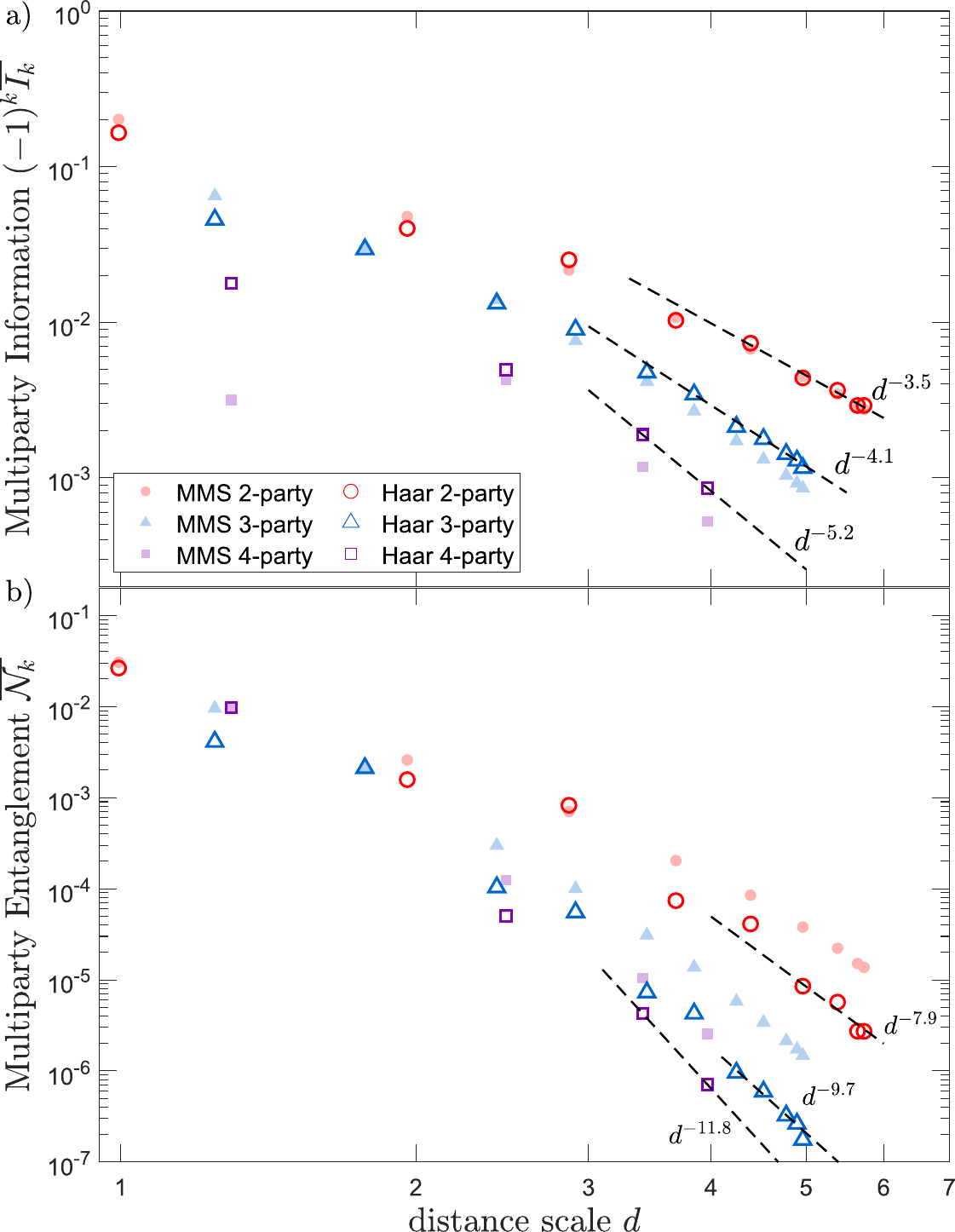}
    \caption{\textbf{Comparison of multiparty correlations in MMS and Haar random circuits.} 
    Data shown for system size $N=18$. 
    \textbf{(a)} Multiparty Mutual Information $ (-1)^k \overline{I_k}$ and \textbf{(b)} Genuine Multiparty Negativity $\overline{\mathcal{N}_k}$ plotted against the distance scale $d$.
    Results for the MMS circuit are represented by filled symbols, while results for the generic Haar-random circuit are represented by open symbols. 
    Dashed lines indicate power-law fits to the Haar data.
    Colors distinguish the number of parties $k$: 2-party (red circles), 3-party (blue triangles), and 4-party (purple squares).}
    \label{fig:mms_haar_comparison}
\end{figure}

Figure~\ref{fig:mms_haar_comparison} compares the distance scaling of $\overline{I_k}$ and $\overline{\mathcal{N}_k}$ for the two ensembles. At $N=18$, the multiparty mutual information exhibits closer quantitative agreement between MMS and Haar than the GMN does. In particular, $\overline{\mathcal{N}_k}$ in the Haar circuit is much lower than in the MMS circuit, consistent with the fact that MMS gates are maximally entangling while a typical Haar gate generates less entanglement on average.

Although the present comparison is limited to a single, relatively small system size, it already illustrates a qualitative contrast that persists throughout our study: mutual-information scaling is comparatively stable, while GMN is substantially more sensitive to circuit details and to finite-size and statistical effects. From a practical perspective, this also clarifies why we focus on the MMS circuit in the main text. The Haar data at accessible sizes can display pronounced even--odd effects, and the overall GMN signal is smaller, so substantially larger system sizes and/or sample counts are required to resolve the long-distance tail with comparable confidence. The stronger signal and smoother finite-size behavior of the MMS circuit therefore provide a cleaner setting for the numerical extraction reported in the main text.

\section{Monogamy constraints on the scaling exponents}\label{app:monogamy_lower_bound}

In this Appendix, we establish a general connection between \emph{monogamy of correlations} and the scaling behavior of multiparty correlation measures at criticality. We first show that any nonnegative $k$-party measure obeying a suitable monogamy relation must satisfy a lower bound on its spatial decay exponent, purely as a consequence of scale invariance. We then apply this general result to two concrete cases studied in this work: bipartite negativity, where a CKW-type monogamy inequality holds rigorously~\cite{CKW_negativity2007}, and multipartite mutual information, where the same structure follows upon assuming sign-definiteness of $\overline{I_k}$. 
While the sign-definiteness of $\overline{I_k}$ cannot be proven in full generality, it has been consistently observed in a range of monitored and chaotic quantum systems, including the negative tripartite mutual information of random Haar circuits~\cite{Pixley2020}, as well as in our numerical data.
It is also worth noting that the monogamy assumptions for multipartite mutual information are naturally motivated by known connections between measurement-induced phase transitions and holographic systems. Hybrid circuit entanglement transitions have been related to holographic entanglement structures~\cite{Li2019_MIPT,Bao2020_MIPT}. In holographic theories, multipartite mutual information is sign-definite and obeys strong monogamy constraints~\cite{Hayden2013_multipartyMI,Mirabi2016_holography_MI_monogamy}. These parallels provide additional theoretical motivation for assuming sign-definiteness of $\overline{I_k}$ in monitored circuits, beyond direct numerical observations reported here and in prior works.

Monogamy of correlations, when combined with scale invariance at criticality, imposes nontrivial constraints on how rapidly multiparty correlation measures can decay with distance. We consider a translationally invariant one-dimensional system and a family of $k$ disjoint subregions $A_1,\ldots,A_k$, each of width $w$ and separated by a distance $r\gg w$. Let $F_k(A_1,\ldots,A_k)$ denote a $k$-party correlation or entanglement measure that is (i) nonnegative, (ii) conformally invariant at criticality,
\begin{equation}
F_k(A_1,\ldots,A_k) \sim \left(\frac{w}{r}\right)^{\alpha_k},
\end{equation}
and (iii) obeys a multipartite monogamy relation
\begin{gather}
    F_k(A_1, A_2, ..., A_{k}) +F_k(A_1', A_2, ..., A_k) \leq F_k(A_1 \cup A_1', A_2, ..., A_k)
\end{gather}
for two disjoint subregions $A_1, A_1'$. To derive a lower bound on $\alpha_k$, we split each region $A_j$ into two equal-width subregions,
\[
A_j = A_j^{(1)} \cup A_j^{(2)}, \qquad 
|A_j^{(1)}| = |A_j^{(2)}| = w/2 ,
\]
with $A_j^{(1)} \cap A_j^{(2)} = \varnothing$.
Applying the monogamy relation to the decomposition $A_1 = A_1^{(1)} \cup A_1^{(2)}$ yields
\begin{align*}
F_k(A_1,A_2,\ldots,A_k)
&\ge F_k\!\bigl(A_1^{(1)},A_2,\ldots,A_k\bigr)
   + F_k\!\bigl(A_1^{(2)},A_2,\ldots,A_k\bigr).
\end{align*}
Since the separation $r \gg w$ is much larger than the internal scale of each region, the two contributions are equal up to corrections of order $O(w/r)$, giving
\begin{align*}
F_k(A_1,A_2,\ldots,A_k)
&\ge 2\,F_k\!\bigl(A_1^{(1)},A_2,\ldots,A_k\bigr)\,(1+O(w/r)).
\end{align*}
Repeating this argument recursively for each region $A_j$, we obtain
\begin{align*}
F_k(A_1,A_2,\ldots,A_k)
&\ge 2^k\,F_k\!\bigl(A_1^{(1)},A_2^{(1)},\ldots,A_k^{(1)}\bigr)\,(1+O(w/r)).
\end{align*}
Finally, using scale invariance at criticality,
\[
F_k\!\bigl(A_1^{(1)},\ldots,A_k^{(1)}\bigr)
= 2^{-\alpha_k} F_k(A_1,\ldots,A_k),
\]
we arrive at
\begin{align*}
F_k(A_1,\ldots,A_k)
&\ge 2^{k-\alpha_k} F_k(A_1,\ldots,A_k)\,(1+O(w/r)),
\end{align*}
which implies the bound
\[
\alpha_k \ge k .
\]

We now apply the general bound to multipartite mutual information. For $k$ subsystems $A_1,\ldots,A_k$, the ensemble-averaged $k$-partite mutual information is defined as
\begin{equation}
\overline{I_k}(A_1,\cdots,A_k)
\equiv \sum_{m=1}^k (-1)^{m-1}
\sum_{1\le i_1<\cdots<i_m\le k}
\overline{S}(A_{i_1}\cup\cdots\cup A_{i_m}),
\end{equation}
where $\overline{S}(\cdot)$ denotes the ensemble-averaged von Neumann entropy. It is convenient to introduce the signed quantity
\begin{equation}
J_k \equiv (-1)^k \overline{I_k} ,
\end{equation}
which is nonnegative whenever $\overline{I_k}$ is alternating-sign definite, i.e.,
$\overline{I_2} \ge 0$, $\overline{I_3} \le 0$, $\overline{I_4} \ge 0$, and so on.

For three subsystems, $J_3 = - \overline{I_3}$ reduces to the tripartite mutual information, and the condition $J_3 \ge 0$ is equivalent to the familiar monogamy of mutual information,
\begin{equation}
I(A,B\cup C) \ge I(A,B) + I(A,C).
\end{equation}
More generally, assuming sign-definiteness of $\overline{I_k}$, the quantities $J_k$ obey a multipartite monogamy relation of the form
\begin{equation}
J_k(A_1\cup A_1',A_2,\ldots,A_k)
\ge J_k(A_1,A_2,\ldots,A_k)
+ J_k(A_1',A_2,\ldots,A_k),
\end{equation}
for disjoint subregions $A_1$ and $A_1'$. Together with nonnegativity and scale invariance at criticality, this shows that $J_k$ satisfies all the requirements of the abstract measure $F_k$ introduced above.

Consequently, applying the general argument yields a lower bound on the scaling exponent of multipartite mutual information,
\begin{equation}
\alpha_k^{\mathrm{MI}} \ge k .
\end{equation}

Finally, we note that in measurement-induced phase transitions the quantities reported are statistical averages over circuit realizations. This does not modify the argument: if the monogamy relation holds for each individual realization, then linearity of the average ensures that the same inequality holds for the averaged quantities. Therefore, the bound $\alpha_k \ge k$ applies equally to the disorder-averaged multipartite mutual information studied in this work.

We now turn to bipartite entanglement as quantified by the negativity. 
For pure states of qubit systems, negativity satisfies a Coffman--Kundu--Wootters (CKW)–type monogamy inequality~\cite{CKW_negativity2007}. 
Specifically, for a pure state $\ket{\psi_{A B_1 \cdots B_m}}$ with all subsystems being qubits, one has
\begin{equation}
\mathcal{N}^2_{A:(B_1\cdots B_m)} \;\ge\; \sum_{j=1}^m \mathcal{N}^2_{A:B_j},
\label{eq:CKW_negativity}
\end{equation}
where $\mathcal{N}_{X:Y}$ denotes the bipartite negativity between subsystems $X$ and $Y$.
This result provides a rigorous monogamy constraint analogous to that satisfied by concurrence.
However, it is important to emphasize that Eq.~\eqref{eq:CKW_negativity} has only been established for qubit systems.
As a consequence, the conformal-invariance-based argument developed above, which relies on subdividing extended regions into smaller subregions, cannot be directly applied in this case.

Instead, we adopt a weaker but fully general argument that relies only on monogamy and asymptotic decay with distance, without invoking conformal geometry.
We consider a one-dimensional chain and a pure state $\ket{\psi_{A B_1 \cdots B_n}}$, where $A$ is a single qubit and $B_r$ denotes the qubit at distance $r$ from $A$.
Suppose the pairwise negativity decays algebraically at long distances,
\begin{equation}
\mathcal{N}_{A:B_r} \sim \mathcal{N}_0\, r^{-\alpha_2}.
\end{equation}
Taking the thermodynamic limit $n \to \infty$, we note that this quantity is automatically bounded:
since $A$ is a single qubit, the bipartite negativity satisfies
$\mathcal{N}_{A:(B_1\cdots B_n)} \le \tfrac{1}{2}$ for all $n$.
Applying the CKW inequality~\eqref{eq:CKW_negativity} then gives
\begin{equation}
\mathcal{N}^2_{A:(B_1\cdots B_n)} \;\ge\; \sum_{r=1}^\infty \mathcal{N}_{A:B_r}^2
\;\sim\; \mathcal{N}_0^2 \sum_{r=1}^\infty r^{-2\alpha_2}.
\end{equation}
The convergence of the series on the right-hand side requires $2\alpha_2>1$, leading to the monogamy constraint
\begin{equation}
\alpha_2 > \tfrac{1}{2}.
\end{equation}
Thus, even without conformal invariance, monogamy alone imposes a nontrivial lower bound on the spatial decay exponent of bipartite negativity.

Finally, we address the fact that in measurement-induced phase transitions the reported negativities are statistical averages over circuit realizations.
Let $k$ label individual realizations occurring with probability $p_k$, and define the vector
\[
\mathbf{v}_k = \bigl(\mathcal{N}^{(k)}_{A:B_1}, \mathcal{N}^{(k)}_{A:B_2}, \ldots, \mathcal{N}^{(k)}_{A:B_m}\bigr).
\]
For each realization, the CKW inequality implies
\begin{equation}
\mathcal{N}^{(k)}_{A:(B_1\cdots B_m)} \;\ge\; \|\mathbf{v}_k\|_2 .
\end{equation}
Averaging over realizations and using the convexity of the Euclidean norm, we obtain
\begin{align}
\overline{\mathcal{N}}_{A:(B_1\cdots B_m)} 
&= \sum_k p_k \mathcal{N}^{(k)}_{A:(B_1\cdots B_m)} \nonumber\\
&\ge \sum_k p_k \|\mathbf{v}_k\|_2
\;\ge\; \Bigl\| \sum_k p_k \mathbf{v}_k \Bigr\|_2 .
\end{align}
Squaring both sides yields the averaged monogamy inequality
\begin{equation}
\overline{\mathcal{N}}_{A:(B_1\cdots B_m)} ^2
\;\ge\; \sum_{j=1}^m \overline{\mathcal{N}}_{A:B_j}^2 .
\end{equation}
Therefore, although our numerics access $\overline{\mathcal{N} }$ rather than $\overline{ \mathcal{N}^2} $, the CKW constraint survives disorder averaging and leads to the same bound $\alpha_2 > 1/2$ for the decay exponent of the disorder-averaged negativity.

\section{Universality of the circuit ensemble} \label{app:ensemble_universality}

We already have, from previous sources that utilized this or similar gate sets~\cite{Sycamore}, that this gate set is universal. However, because the gate set does not initially contain an identity, it is nontrivial to prove that a brickwork circuit of gates drawn only from this set is capable of approximating all $N$-site unitaries.

We start with one specific unitary in our limited gate set,
\begin{align}
    U^{xx} &= M \; (R^x \otimes R^x)\notag\\
    &=\frac{1}{\sqrt{2}}\left(R^x \otimes R^x +iR^{x\dagger} \otimes R^{x\dagger}\right)
\end{align}
Here we use superscripts to denote axes of rotation, e.g $R^x = R(\hat{x}), R^y = R(\hat{y})$. $U^{xx}$ squares to the identity, up to an overall phase. Moreover, neighboring $U^{xx}$ gates that share a site in common still commute, since their individual components commute. Therefore, four brickwork layers of $U^{xx}$ (two even and two odd layers) is the same as applying $U^{xx}$ twice to every even and odd bond, which is the same as the identity on the whole system. 

Our strategy will be as follows: now that we have a way to implement the identity through a brickwork of gates, we will fill the circuit with copies of this identity, and then create single-site and two-site unitaries by modifying each copy. Without loss of generality, we will assume that the brickwork layers making up each identity starts with an even layer on the bottom and ends with an odd layer on the top. We will start by creating single-site unitaries. If we replace $U^{xx}$ with $U^{x\alpha}$ on the bottom layer, that is equivalent to applying $R^{x\dagger} R^\alpha$ to the affected qubit, and the set of such operations is universal to SU(2). Therefore, we can create arbitrary single-site unitary gates.

Now let's replace one of the $U^{xx}$ gates on the top layer with $U^{yy}$. This is equivalent to applying $U^{yy} U^{xx\dagger}$ to the affected qubits. This looks like
\begin{align}
    U^{yy} U^{xx\dagger} &= \frac{1}{2}\left(I \otimes I -i X \otimes X\right) \left( R^y R^{x\dagger} \otimes R^y R^{x\dagger}\right)\left(I \otimes I +i X \otimes X\right)\notag \\
    &= \frac{1}{2}\left(R^y R^{x \dagger} \otimes R^y R^{x \dagger}\right) \left(I \otimes I -iY\otimes Y\right) \left(I \otimes I +i X \otimes X\right)\notag \\
    &= \frac{1}{2}\left(R^y R^{x \dagger} \otimes R^y R^{x \dagger}\right) \left(I \otimes X\right) (\text{iSWAP})\left(I\otimes X\right)
\end{align}
so is equivalent to iSWAP, up to single-site unitaries. Since iSWAP and single-site unitaries are universal to SU(4)~\cite{Echternach2001}, we can generate all two-site unitaries on all odd neighbors $(i,i+1),i=2k+1$ of the circuit (odd neighbors because of our assumption that the identity brickworks are ending on an odd layer).

Next, we implement two-site unitaries on even neighbors. Suppose, for an even pair of sites $(i,i+1),i=2k$, we replaced the $U^{xx}_{(i,i+1)}$ gate acting on that pair in the highest even layer (the 2nd highest layer in the identity) with a $U^{yy}_{(i,i+1)}$ gate. This is equivalent to applying the sequence
\begin{gather}
    U^{xx}_{i-1,i}U^{xx}_{i+1,i+2} U^{yy}_{i,i+1} (U^{xx}_{i,i+1})^\dagger (U^{xx}_{i-1,i})^\dagger (U^{xx}_{i+1,i+2})^\dagger
\end{gather}
to the top of the identity brickwork. But since $(i-1,i)$ and $(i+1,i+2)$ are both odd pairs, we already know how to implement $U^{xx}$ on those pairs (note that $U^{xx}_{i,i+1}$ is equal to its conjugate $(U^{xx}_{i,j})^\dagger$, up to a constant phase). So, if we implement $U^{xx}_{i-1,i} U^{xx}_{i+1,i+2}$ before this modification, and $(U^{xx}_{i-1,i})^\dagger (U^{xx}_{i+1,i+2})^\dagger$ after this modification, we have effectively implemented $U^{yy}_{i,i+1} (U^{xx \dagger}_{i,i+1})$, which we already know is equivalent to an iSWAP under single-site unitaries, and therefore can be used to generate all other two-site unitaries on the pair $(i,i+1)$. Hence, we can generate two-site unitaries on both even and odd neighbors, and therefore we can generate any unitary circuit.

\section{Comparison with deeper circuit} \label{app:comparison_deeper_circuit}

\begin{figure}[hbt!]
    \centering
    \includegraphics[width=0.5\textwidth]{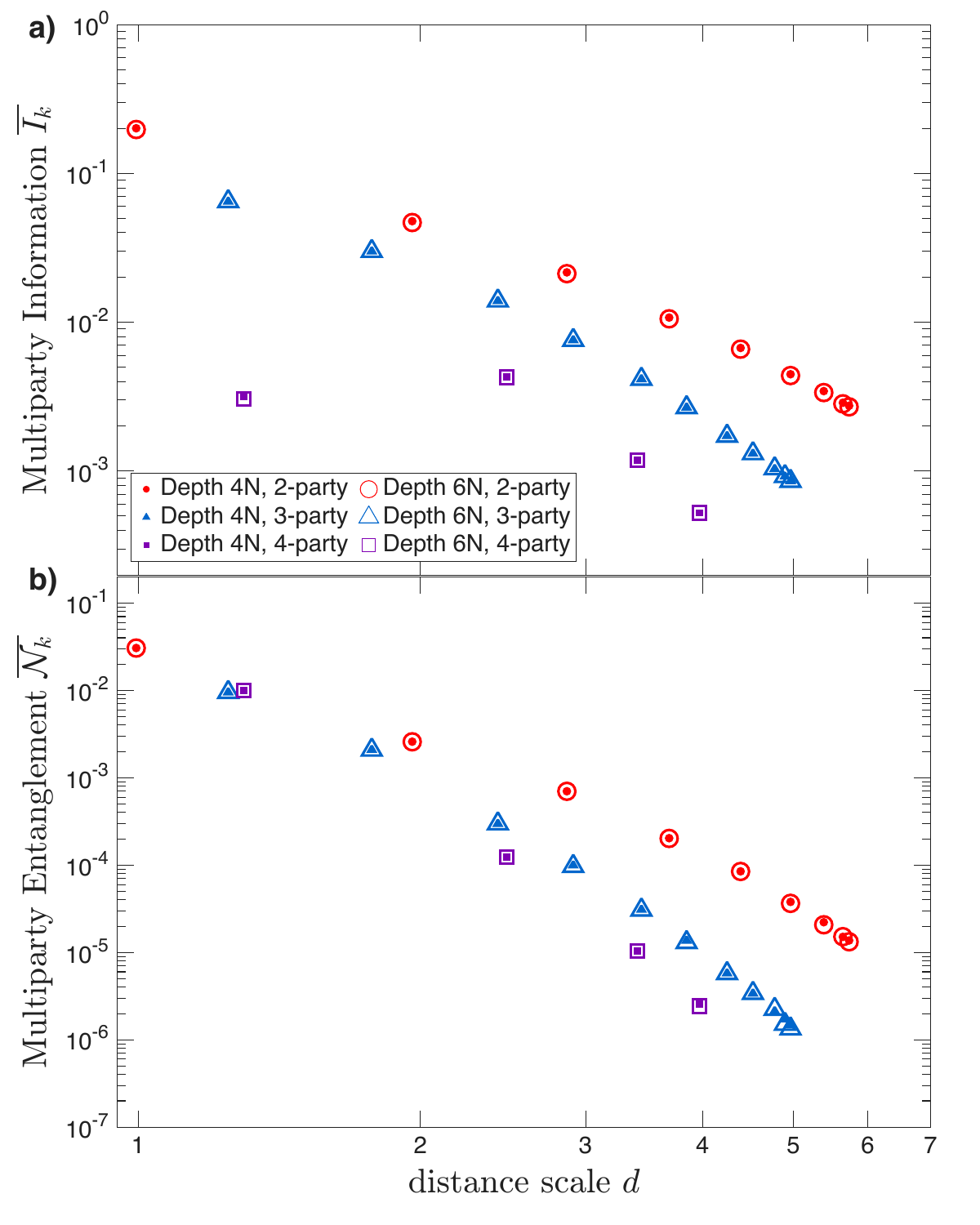}
    \caption{\textbf{Convergence of multiparty correlations with circuit depth.} Comparison of entanglement metrics for $N=18$ at standard circuit depth $D=4N$ (small filled symbols) and extended depth $D=6N$ (large open symbols). 
    \textbf{(a)} Multiparty Mutual Information $\overline{I_k}$ and \textbf{(b)} Genuine Multiparty Negativity $\overline{\mathcal{N}_k}$ as a function of the distance scale $d$. 
    Different colors denote the number of parties $k$: 2-party (red circles), 3-party (blue triangles), and 4-party (purple squares). 
    The strong overlap between the two datasets indicates that the system has reached a steady state at $D=4N$.}
    \label{fig:depth_comparison}
\end{figure}

In this section, we verify that the circuit depth $D=4N$ ($2N$ unitary layers $+$ $2N$ measurement layers) employed in the main text is sufficient for the system to reach its non-equilibrium steady state. To demonstrate this, we compare our main results against benchmark simulations performed at a larger depth of $D=6N$ for a system size of $N=18$.

As illustrated in Fig.~\ref{fig:depth_comparison}, the data for both Multiparty Mutual Information ($\overline{I_k}$) and Genuine Multiparty Negativity ($\overline{\mathcal{N}_k}$) at $D=4N$ (filled symbols) are statistically indistinguishable from those obtained at $D=6N$ (open symbols) across all distance scales. This strong overlap confirms that transient dynamics have decayed and that the correlations have saturated, justifying our use of $D=4N$ for the analysis presented in the main text.

\section{Statistical sample sizes and positive event counts}~\label{app:sample_sizes}

Table~\ref{tab:count_count} lists the total number of circuit realizations and the resulting counts of positive (non-zero) events for each measure (GMN and multiparty mutual information) across system sizes $N=18$ to $26$. These values provide the statistical basis for the spatial scaling of multiparty entanglement shown in Fig.~\ref{fig:ENT_dist_scaling} and the decay exponents extracted in Table~\ref{tab:entanglement_exponents}.
For three-party metrics, ``A'' denotes asymmetric configurations with unequal subregion spacings $\{i, i+x, i+2x+1\}$. The number of evaluated reduced density matrices for each configuration is proportional to the circuit count and system size, adjusted for symmetries at maximal separations.

\begin{table}[h]
\centering
\begin{tabular}{|l|l|c||c|c|c|c|c|c|c|c|c|c|c|c|c|}
\hline
\textbf{N} & \textbf{Quantity} & \textbf{Circuits} &x=1&x=2&x=3&x=4&x=5&x=6&x=7&x=8&x=9&x=10&x=11&x=12&x=13\\ \hline
18&2-GMN&6688716&29M&4805K&1675K&645K&318K&169K&110K&81.5K&75.6K&&&&\\ \hline
&3-GMN&&14.9M&906K&148K&38.8K&17.8K&13.5K&&&&&&&\\ \hline
&3-GMN-A&&5867K&525K&108K&36K&21.2K&&&&&&&&\\ \hline
&4-GMN&&14.6M&545K&76.6K&24.6K&&&&&&&&&\\ \hline
&2-MI&&65.5M&59.2M&56.9M&55.4M&54.9M&54.6M&54.6M&54.5M&54.5M&&&&\\ \hline
&3-MI&&4964K&2372K&1920K&1761K&1732K&1739K&&&&&&&\\ \hline
&3-MI-A&&5451K&3118K&2655K&2520K&2504K&&&&&&&&\\ \hline
&4-MI&&20.1M&26.5M&25.1M&23.7M&&&&&&&&&\\ \hline
20&2-GMN&10310506&46.3M&7530K&2552K&949K&448K&221K&131K&85.4K&67.6K&61.7K&&&\\ \hline
&3-GMN&&25.4M&1448K&212K&46.9K&17K&9683&&&&&&&\\ \hline
&3-GMN-A&&9945K&816K&148K&40.2K&17.8K&&&&&&&&\\ \hline
&4-GMN&&24.3M&782K&85.1K&19.4K&12.2K&&&&&&&&\\ \hline
&2-MI&&108M&97.6M&93.7M&91.2M&93.4M&93.6M&93.5M&93.3M&92.6M&89.9M&&&\\ \hline
&3-MI&&8539K&4029K&3252K&3008K&2975K&3004K&&&&&&&\\ \hline
&3-MI-A&&9512K&5507K&4738K&4529K&4523K&&&&&&&&\\ \hline
&4-MI&&33.2M&43.6M&41.2M&38.8M&38.3M&&&&&&&&\\ \hline
22&2-GMN&15444254&76.7M&12.2M&4091K&1474K&675K&318K&179K&106K&75.1K&58.3K&54.8K&&\\ \hline
&3-GMN&&38.3M&2086K&284K&55.7K&16.9K&7578&5283&&&&&&\\ \hline
&3-GMN-A&&12.7M&987K&162K&38.3K&13.9K&7550&&&&&&&\\ \hline
&4-GMN&&37.5M&1081K&99.3K&16.9K&7145&&&&&&&&\\ \hline
&2-MI&&180M&163M&157M&152M&151M&150M&150M&150M&150M&150M&150M&&\\ \hline
&3-MI&&13.8M&6716K&5542K&5166K&5138K&5216K&5255K&&&&&&\\ \hline
&3-MI-A&&13.2M&7715K&6694K&6443K&6484K&6561K&&&&&&&\\ \hline
&4-MI&&55.5M&72.5M&68.6M&64.6M&63.2M&&&&&&&&\\ \hline
24&2-GMN&26275721&140M&22M&7278K&2573K&1155K&523K&282K&158K&103K&72K&59.5K&33.2K&\\ \hline
&3-GMN&&72.6M&3824K&493K&89.7K&23.8K&8701&4864&1908&&&&&\\ \hline
&3-GMN-A&&30.9M&2301K&354K&75.6K&23.3K&10.4K&5953&&&&&&\\ \hline
&4-GMN&&71.5M&1913K&153K&20.8K&6346&1820&&&&&&&\\ \hline
&2-MI&&337M&305M&293M&285M&282M&281M&281M&280M&280M&280M&280M&171M&\\ \hline
&3-MI&&25.8M&12.7M&10.5M&9866K&9860K&10M&10.2M&4951K&&&&&\\ \hline
&3-MI-A&&31.8M&18.7M&16.3M&15.8M&16M&16.3M&14.4M&&&&&&\\ \hline
&4-MI&&104M&135M&128M&120M&117M&48.4M&&&&&&&\\ \hline
26&2-GMN&165821&952K&149K&48.9K&16.9K&7389&3230&1819&912&554&386&264&193&144\\ \hline
&3-GMN&&491K&25.7K&3184&486&143&30&18&12&&&&&\\ \hline
&3-GMN-A&&258K&18.9K&2722&514&148&61&35&&&&&&\\ \hline
&4-GMN&&482K&12.6K&922&96&26&10&&&&&&&\\ \hline
&2-MI&&2301K&2083K&1999K&1947K&1929K&1920K&1917K&1915K&1913K&1912K&1911K&1911K&1348K\\ \hline
&3-MI&&177K&86.9K&72.6K&68.7K&69K&70.9K&71.9K&73K&&&&&\\ \hline
&3-MI-A&&271K&160K&141K&137K&140K&142K&145K&&&&&&\\ \hline
&4-MI&&709K&922K&870K&818K&796K&785K&&&&&&&\\ \hline
\end{tabular}
\caption{\textbf{Number of positive counts detected for each type of entanglement.} The total number of RDMs evaluated for each count type is equal to the number of circuits times $N$, unless the number of parties divides $N$ and $x$ is at its maximal value for the specific entanglement type, in which case the number of RDMs must be divided by the number of parties. Thousands and millions are abbreviated to K and M respectively. ``3-GMN-A'' and ``3-MI-A'' indicate three-party entanglement/MI among subregions of the form $\{i,i+x,i+2x+1\}$ or $\{i,i+x+1,i+2x+1\}$.}
\label{tab:count_count}
\end{table}

\begin{table}[h]
\centering
\begin{tabular}{|l|c|c|c|c|}
\hline
\textbf{p} & \textbf{N=12} & \textbf{N=16} & \textbf{N=20} & \textbf{N=24} \\ \hline
0.11&240K&&& \\ \hline 
0.12&222K&177K&238K& \\ \hline 
0.125&208K&152K&266K& \\ \hline 
0.13&214K&202K&467K&210K \\ \hline 
0.135&207K&252K&296K&365K \\ \hline 
0.14&343K&152K&298K&430K \\ \hline 
0.145&225K&152K&293K&424K \\ \hline 
0.1475&240K&200K&150K&150K \\ \hline 
0.15&239K&399K&253K&428K \\ \hline 
0.1525&224K&300K&377K&309K \\ \hline 
0.155&239K&162K&155K&262K \\ \hline 
0.1575&234K&300K&200K&222K \\ \hline 
0.16&242K&173K&318K&197K \\ \hline 
0.1625&205K&150K&200K&164K \\ \hline 
0.165&240K&431K&282K&256K \\ \hline 
0.1675&242K&300K&200K&150K \\ \hline 
0.17&240K&199K&271K&279K \\ \hline 
0.1725&150K&200K&150K&150K \\ \hline 
0.175&160K&237K&165K&179K \\ \hline 
0.18&528K&183K&220K&255K \\ \hline 
0.185&1282K&226K&166K&317K \\ \hline 
0.19&1270K&268K&197K&300K \\ \hline 
0.195&201K&190K&170K& \\ \hline 
0.2&201K&190K&170K& \\ \hline 
\end{tabular}
\caption{\textbf{Number of circuits simulated for TMI measurements.} The total number of circuits simulated for each pair of lengths $N$ and measurment probability $p$ during calculations of ensemble-averaged TMI, $\overline{I_3}(A,B,C)$, from Section~\ref{sec:critical_point}.}
\label{tab:TMI_circuit_count}
\end{table}

\end{appendices}

\end{document}